\documentclass[twocolumn,american,nofootinbib,preprintnumbers]{revtex4}
\usepackage{ae}
\usepackage{aecompl}
\usepackage[T1]{fontenc}
\usepackage[latin1]{inputenc}
\usepackage{amsmath}
\usepackage{graphicx}
\usepackage{amssymb}

\makeatletter
\usepackage{psfrag}

\usepackage{babel}
\makeatother
\begin{document}

\preprint{YITP-08-41}

\title{Reheating-volume measure for random-walk inflation}

\author{Sergei Winitzki$^{1,2}$}

\affiliation{$^{1}$Department of Physics, Ludwig-Maximilians University, Munich,
Germany}

\affiliation{$^{2}$Yukawa Institute of Theoretical Physics, Kyoto University,
Kyoto, Japan }

\begin{abstract}
The recently proposed {}``reheating-volume'' (RV) measure promises
to solve the long-standing problem of extracting probabilistic predictions
from cosmological {}``multiverse'' scenarios involving eternal inflation.
I give a detailed description of the new measure and its applications
to generic models of eternal inflation of random-walk type. For those
models I derive a general formula for RV-regulated probability distributions
that is suitable for numerical computations. I show that the results
of the RV cutoff in random-walk type models are always gauge-invariant
and independent of the initial conditions at the beginning of inflation.
In a toy model where equal-time cutoffs lead to the {}``youngness
paradox,'' the RV cutoff yields unbiased results that are distinct
from previously proposed measures. 
\end{abstract}
\maketitle

\section{Introduction and motivation\label{sec:Introduction-and-motivation}}

 It was realized in recent years that in many cosmological scenarios
the fundamental theory does not predict with certainty the values
of observable cosmological parameters, such as the effective cosmological
constant and the masses of elementary particles. This is the case
for the {}``landscape of string theory''~\cite{Bousso:2000xa,Susskind:2003kw,Douglas:2003um}
(see also the {}``recycling universe''~\cite{Garriga:1997ef})
and for models of inflation driven by a scalar field (see e.g.~\cite{Garcia-Bellido:1993wn,Garcia-Bellido:1994vz}
for early work). A common feature of these cosmological models is
the presence of \emph{eternal inflation}, i.e.~the absence of a global
end to inflation in the entire spacetime (see Refs.~\cite{Linde:1993xx,Guth:2000ka,Winitzki:2006rn}
for reviews). Eternal inflation gives rise to infinitely many causally
disconnected regions of the spacetime where the cosmological observables
may have significantly different values. Hence the program outlined
in the early works~\cite{Garcia-Bellido:1994ci,Vilenkin:1994ua,Vilenkin:1995yd}
was to obtain the probability distribution of the cosmological parameters
as measured by an observer randomly located in the spacetime. The
main diffuculty in obtaining such probability distributions is due
to the infinite volume of regions where an observer may be located. 

Since the spacetime during inflationary evolution is cold and empty,
observers may appear only after reheating. The standard cosmology
after reheating is tightly constrained by current experimental knowledge.
Hence, the average number of observers produced in any freshly-reheated
spatial domain is a function of cosmological parameters in that domain.
Calculating that function is, in principle, a well-defined astrophysical
problem that does not involve any infinities. Therefore we focus on
the problem of obtaining the probability distribution of cosmological
observables at reheating.

\begin{figure}
\begin{centering}\psfrag{x}{$x$}\psfrag{t}{$t$}\includegraphics[width=0.85\columnwidth,keepaspectratio]{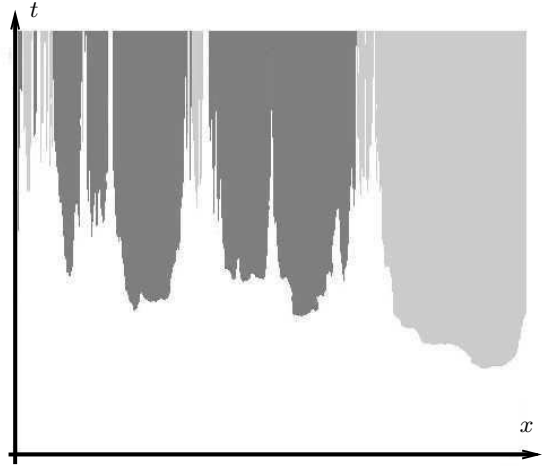}\par\end{centering}

\caption{A 1+1-dimensional slice of the spacetime in an eternally inflating
universe (numerical simulation in Ref.~\cite{Vanchurin:1999iv}).
Shades of different color represent different regions where reheating
took place. The reheating surface is the line separating the white
(inflating) domain and the shaded domains.\label{cap:reheating-surface-1} }
\end{figure}

The set of all spacetime points where reheating takes place is a spacelike
three-dimensional hypersurface~\cite{Borde:1993xh,Vilenkin:1995yd,Creminelli:2008es}
called the {}``reheating surface.'' The hallmark feature of eternal
inflation is that a \emph{finite}, initially inflating spatial 3-volume
typically gives rise to a reheating surface having an \emph{infinite}
3-volume (see Fig.~\ref{cap:reheating-surface-1}). The geometry
and topology of the reheating surface is quite complicated. For instance,
the reheating surface contains infinitely many future-directed spikes
around never-thermalizing comoving worldlines called {}``eternally
inflating geodesics''~\cite{Winitzki:2001np,Winitzki:2005fy,Vanchurin:2006xp}.
It is known that the set of {}``spikes'' has a well-defined fractal
dimension that can be computed in the stochastic approach~\cite{Winitzki:2001np}.
Since the reheating surface is a highly inhomogeneous, noncompact
3-manifold without any symmetries, a {}``random location'' on such
a surface is mathematically ill-defined. This feature of eternal inflation
is at the root of the technical and conceptual difficulties known
collectively as the {}``measure problem'' (see Refs.~\cite{Guth:2000ka,Aguirre:2006ak,Winitzki:2006rn,Vilenkin:2006xv,Guth:2007ng,Linde:2007nm}
for reviews).

To visualize the measure problem, it is convenient to consider an
initial inflating spacelike region $S$ of horizon size (an {}``$H$-region'')
and the portion $R\equiv R(S)$ of the reheating surface that corresponds
to the comoving future of $S$. If the 3-volume of $R$ were finite,
the volume-weighted average of any observable quantity $Q$ at reheating
would be defined simply by averaging $Q$ over $R$,\begin{equation}
\left\langle Q\right\rangle \equiv\frac{\int_{R}Q\sqrt{\gamma}d^{3}x}{\int_{R}\sqrt{\gamma}d^{3}x},\label{eq:Q average}\end{equation}
where $\gamma$ is the induced metric on the 3-surface $R$. This
would have been the natural prescription for the observer-based average
of $Q$; all higher moments of the distribution of $Q$, such as $\left\langle Q^{2}\right\rangle $,
$\left\langle Q^{3}\right\rangle $, etc., would have been well-defined
as well. However, in the presence of eternal inflation%
\footnote{Various equivalent conditions for the presence of eternal inflation
were examined in more detail in Refs.~\cite{Winitzki:2001np,Winitzki:2005ya}
and \cite{Creminelli:2008es}. Here I adopt the condition that $X(\phi)$
is nonzero for all $\phi$ in the inflating range.%
} the 3-volume of $R$ is infinite with a nonzero probability $X(\phi_{0})$,
where $\phi=\phi_{0}$ is the initial value of the inflaton field
at $S$. The function $X(\phi_{0})$ has been computed in slow-roll
inflationary models~\cite{Winitzki:2001np} where typically $X(\phi_{0})\approx1$
for $\phi_{0}$ not too close to reheating. In other words, the volume
of $R$ is infinite with a probability close to 1. In that case, the
straightforward average~(\ref{eq:Q average}) of a fluctuating quantity
$Q(x)$ over $R$ is mathematically undefined since $\int_{R}\sqrt{\gamma}d^{3}x=\infty$
and $\int_{R}Q\sqrt{\gamma}d^{3}x=\infty$. 

The average $\left\langle Q\right\rangle $ can be computed only after
imposing a volume cutoff on the reheating surface, making its volume
finite in a controlled way. What has become known in cosmology as
the {}``measure problem'' is the difficulty of coming up with a
physically motivated cutoff prescription (informally called a {}``measure'')
that makes volume averages $\left\langle Q\right\rangle $ well-defined. 

Volume cutoffs are usually implemented by restricting the infinite
reheating domain $R$ to a large but finite subdomain having a volume
$\mathcal{V}$. Then one defines the {}``regularized'' distribution
$p(Q|\mathcal{V})$ of an observable $Q$ by gathering statistics
about the values of $Q$ over the finite volume $\mathcal{V}$. More
precisely, $p(Q|\mathcal{V})\mathcal{V}dQ$ is the 3-volume of regions
(within the finite domain $\mathcal{V}$) where the observable $Q$
has values in the interval $\left[Q,Q+dQ\right]$. The final probability
distribution $p(Q)$ is then defined as \begin{equation}
p(Q)\equiv\lim_{\mathcal{V}\rightarrow\infty}p(Q|\mathcal{V}),\end{equation}
provided that the limit exists. 

Several cutoffs have been proposed in the literature, differing in
the choice of the compact subset $\mathcal{V}$ and in the way $\mathcal{V}$
approaches infinity. It has been found early on (e.g.~\cite{Linde:1993xx,Vilenkin:1995yd})
that probability distributions, such as $p(Q)$, depend sensitively
on the choice of the cutoff. This is the root of the measure problem.
Since a {}``natural'' mathematically consistent definition of the
measure is absent, one judges a cutoff prescription viable if its
predictions are not obviously pathological. Possible pathologies include
the dependence on choice of spacetime coordinates~\cite{Winitzki:1995pg,Linde:1995uf},
the {}``youngness paradox''~\cite{Linde:1994gy,Vilenkin:1998kr},
and the {}``Boltzmann brain'' problem~\cite{Dyson:2002pf,Albrecht:2004ke,Page:2006dt,Linde:2006nw,Vilenkin:2006qg,Page:2006ys,Bousso:2007nd,Gott:2008ii}.

The presently viable cutoff proposals fall into two rough classes
that may be designated as {}``worldline-based'' and {}``volume-based''
measures (a more fine-grained classification of measure proposals
can be found in Refs.~\cite{Aguirre:2006ak,Vanchurin:2006xp}). The
{}``worldline'' or the {}``holographic'' measure~\cite{Bousso:2006ev,Bousso:2006ge}
avoids considering the infinite total 3-volume of the reheating surface
in the entire spacetime. Instead it focuses only on the reheated 3-volume
of one $H$-region surrounding a \emph{single} randomly chosen comoving
worldline. This measure, by construction, is sensitive to the initial
conditions at the location where the worldline starts and is essentially
equivalent to performing calculations with the comoving-volume probability
distribution. Proponents of the {}``holographic'' measure have argued
that the infinite reheating surface cannot be considered because the
spacetime beyond one $H$-region is not adequately described by semiclassical
gravity~\cite{Bousso:2006ge}. However, the semiclassical approximation
was recently shown to be valid in a large class of inflationary models~\cite{Tolley:2008na}.
In my view, an attempt to count the total volume of the reheating
surface corresponds more closely to the goal of obtaining the probability
distribution of observables in the entire universe, as measured by
a {}``typical'' observer (see Refs.~\cite{Pogosian:2006fx,Maor:2007wq,Garriga:2007wz,Hartle:2007zv,Bousso:2007nd}
for recent discussions of {}``typicality'' and accompanying issues).
The sensitive dependence of {}``holographic'' proposals on the conditions
at the beginning of inflation also appears to be undesirable. Volume-based
proposals are insensitive to the initial conditions because the 3-volume
of the universe is, in a certain well-defined sense, dominated by
regions that spent a long time in the inflationary regime.%
\footnote{It has been noted that 3-volume is a coordinate-dependent quantity,
and hence statements involving 3-volume need to be formulated with
care~\cite{Gratton:2005bi}. Indeed there exist time foliations where
the 3-volume of inflationary space does not grow with time. The issues
of coordinate dependence were analyzed in Ref.~\cite{Winitzki:2005ya}.%
} Existing volume-based proposals include the equal-time cutoff~\cite{Garcia-Bellido:1993wn,Garcia-Bellido:1994ci,Garcia-Bellido:1994vz},
the {}``spherical cutoff''~\cite{Vilenkin:1998kr}, the {}``comoving
horizon cutoff''~\cite{Garriga:2005av,Easther:2005wi,Vanchurin:2006qp},
the {}``stationary measure''~\cite{Linde:2007nm,Clifton:2007bn},
the {}``no-boundary'' measure with volume weighting~\cite{Hawking:2007zz,Hartle:2007gi,Hawking:2007vf,Hartle:2008ng},
the {}``pseudo-comoving'' measure~\cite{Linde:2006nw,DeSimone:2008bq},
and the most recently proposed {}``reheating-volume'' (RV) measure~\cite{Winitzki:2008yb}. 

The focus of this article is a more detailed study of the RV measure
in the context of random-walk eternal inflation. As a typical generic
model I choose a scenario where inflation is driven by the potential
$V(\phi)$ of a minimally coupled scalar field $\phi$. In this model,
there exists a range of $\phi$ where large quantum fluctuations dominate
over the deterministic slow-roll evolution, which gives rise to eternal
self-reproduction of inflationary domains. I extensively use the stochastic
approach to inflation, which is based on the Fokker-Planck or {}``diffusion''
equations (see Ref.~\cite{Winitzki:2006rn} for a pedagogical review).
The results can be straightforwardly generalized to multiple-field
or non-slow-roll models are straightforward since the Fokker-Planck
formalism is already developed in those contexts~\cite{Helmer:2006tz,Tolley:2008na}.
Applications of the RV measure to {}``landscape'' scenarios will
be considered elsewhere. 

An attractive feature of the RV measure is that its construction lacks
extraneous geometric elements that could introduce a bias. An example
of a biased measure is the equal-time cutoff where one considers the
subdomain of the reheating surface to the past of a hypersurface of
fixed proper time $t=t_{c}$, subsequently letting $t_{c}\rightarrow\infty$.
It is well known that the volume-weighted distribution of observables
within a hypersurface of equal proper time is strongly dominated by
regions where inflation ended very recently. A time delay $\delta t$
in the onset of reheating due to a rare quantum fluctuation is overwhelmingly
rewarded by an additional volume expansion factor $\propto\exp[3H_{\max}\delta t]$,
where $H_{\max}$ is roughly the highest Hubble rate accessible to
the inflaton. This is the essence of the so-called youngness paradox
that seems unavoidable in an equal-time cutoff (see Refs.~\cite{Tegmark:2004qd}
and \cite{Bousso:2007nd} for recent discussions). 

Moreover, the results of the equal-time cutoff are sensitive to the
choice of the time coordinate ({}``time gauge''). For instance,
the proper time can be replaced by the family of time gauges labeled
by a constant $\alpha$,~\cite{Winitzki:1995pg}\begin{equation}
t_{(\alpha)}\equiv\int^{t}H^{\alpha}dt,\end{equation}
which interpolate between the proper time ($\alpha=0$, $t_{(0)}\equiv t$)
and the $e$-folding time ($\alpha=1$, $t_{(1)}=\ln a$). It has
been shown that the results of the equal-time cutoff depend sensitively
on the value of $\alpha$, and that no {}``correct'' value of $\alpha$
could be specified so as to remove the bias~\cite{Winitzki:2005ya}.
Since the time coordinate is an arbitrary label in the spacetime,
we may impose the requirement that a viable measure prescription be
invariant with respect to choosing even more general time gauges,
such as\begin{equation}
\tau\equiv\int^{t}T(\phi)dt,\label{eq:general time gauge}\end{equation}
where $T(\phi)>0$ is an arbitrary function of the inflaton field
(and possibly of other fields), and the integration is performed along
comoving worldlines $x^{1,2,3}=\mbox{const}$.

The {}``spherical cutoff''~\cite{Vilenkin:1998kr} and the {}``stationary
measure''~\cite{Linde:2007nm} prescriptions were motivated by the
need to remove the bias inherent in the equal-time cutoff. In particular,
the spherical cutoff selects as a compact subset $\mathcal{V}$ the
interior of a large sphere drawn within the reheating surface $R$
around a randomly chosen center. The spherical cutoff is manifestly
gauge-invariant since its construction uses only the intrinsically
defined 3-volume of the reheating surface rather than the spacetime
coordinates $\left(t,x\right)$. Some results were obtained in the
spherical cutoff using numerical simulations~\cite{Vanchurin:1999iv}.
A disadvantage of the spherical cutoff is that its direct implementation
requires one to perform costly numerical simulations of random-walk
inflation on a spacetime grid, for instance, using the techniques
of Refs.~\cite{Linde:1993xx,Vanchurin:1999iv,Winitzki:1999ve}. Instead,
one would prefer to obtain a generally valid analytic formula for
the probability distribution of cosmological observables. For instance,
one could ask whether the results of the spherical cutoff depend in
an essential way on the spherical shape of the region, on the position
of the center of the sphere, and on the initial conditions. Satisfactory
answers to these questions (in the negative) were obtained in Refs.~\cite{Vilenkin:1998kr,Vanchurin:1999iv}
in some tractable cases where results could be obtained analytically.
However, it is difficult to analyze these questions in full generality
since one lacks a general analytic formula for the probability distribution
in the spherical cutoff. 

The RV measure is similar in spirit to the spherical cutoff because
the RV cutoff uses only the intrinsic geometrical information defined
by the reheating surface. It can be argued that the RV cutoff is {}``more
natural'' than other cutoffs in that it selects a finite portion
$\mathcal{V}$ of the reheating surface without using artificial constant-time
hypersurfaces, spheres, worldlines, or any other \emph{}extraneous
geometrical data. Instead, the selection of $\mathcal{V}$ in the
RV cutoff is performed using a certain well-defined selection of subensemble
in the probability space, which is determined by the stochastic evolution
itself.

The central concept in the RV cutoff is the {}``finitely produced
volume.'' The basic idea is that there is always a nonzero probability
that a given initial $H$-region $S$ \emph{does} \emph{not} give
rise to an infinite reheating surface in its comoving future. For
instance, it is possible that by a rare coincidence the inflaton field
$\phi$ rolls towards reheating at approximately the same time everywhere
in $S$. Moreover, there is a nonzero (if small) probability $\rho(\mathcal{V})d\mathcal{V}$
that the total volume $\mbox{Vol}(R)$ of the reheating surface $R$
to the future of $S$ belongs to a given interval $\left[\mathcal{V},\mathcal{V}+d\mathcal{V}\right]$,
\begin{equation}
\rho(\mathcal{V})\equiv\lim_{d\mathcal{V}\rightarrow0}\frac{\mbox{Prob}\left\{ \mbox{Vol}(R)\in\left[\mathcal{V},\mathcal{V}+d\mathcal{V}\right]\right\} }{d\mathcal{V}}.\label{eq:rho def}\end{equation}
I call $\rho(\mathcal{V})$ the {}``finitely produced volume distribution.''
This distribution is nontrivial because the probability of the event
$\mbox{Vol}(R)<\infty$ is nonzero, if small, for any given (non-reheated)
initial region $S$. The distribution $\rho(\mathcal{V})$ is, by
construction, normalized to that probability:\begin{equation}
\int_{0}^{\infty}\rho(\mathcal{V})d\mathcal{V}=\mbox{Prob}\left\{ \mbox{Vol}(R)<\infty\right\} <1.\label{eq:rho V normalization}\end{equation}

The RV cutoff consists of a selection of a certain ensemble $E_{\mathcal{V}}$
of the histories that produce a total reheated volume equal to a given
value $\mathcal{V}$ starting from an initial $H$-region. In the
limit of large $\mathcal{V}$, the ensemble $E_{\mathcal{V}}$ consists
of $H$-regions that evolve {}``almost'' to the regime of eternal
inflation. Thus, heuristically one can expect that the ensemble $E_{\mathcal{V}}$
provides a representative sample of the infinite reheating surface.%
\footnote{Of course, this heuristic statement cannot be made rigorous since
there exists no natural measure on the infinite reheating surface.
We use this statement merely as an additional motivation for considering
the RV measure.%
} Given the ensemble $E_{\mathcal{V}}$, one can determine the volume-weighted
probability distribution $p(Q|{\cal E}_{\mathcal{V}})$ of a cosmological
parameter $Q$ by ordinary sampling of the values of $Q$ throughout
the finite volume $\mathcal{V}$. Finally, the probability distribution
$p(Q)$ is defined as the limit of $p(Q|{\cal E}_{\mathcal{V}})$
at $\mathcal{V}\rightarrow\infty$, provided that the limit exists.

To clarify the construction of the ensemble $E_{\mathcal{V}}$, it
is helpful to begin by considering the distribution $\rho(\mathcal{V})$
in a model that does \emph{not} permit eternal inflation. In that
case, the volume of the reheating surface is finite with probability
1, so the distribution $\rho(\mathcal{V})$ is an ordinary probability
distribution normalized to unity. In that context, the distribution
$\rho(\mathcal{V})$ was introduced in the recent work~\cite{Creminelli:2008es}
where the authors considered a family of inflationary models parameterized
by a number $\Omega$, such that eternal inflation is impossible in
models where $\Omega>1$. It was then found by a direct calculation
that all the moments of the distribution $\rho(\mathcal{V})$ diverge
at the value $\Omega=1$ where the possibility of eternal inflation
is first switched on. One can show that the \emph{finitely produced}
distribution $\rho(\mathcal{V})$ for $\Omega<1$ is again well-behaved
and has finite moments (see Sec.~\ref{sub:Asymptotics-of-rho}).
This FPRV distribution $\rho(\mathcal{V})$ is the formal foundation
of the RV cutoff. It is worth emphasizing that the RV cutoff \emph{}does
\emph{}not regulate the volume of the reheating surface by modifying
the dynamics of a given inflationary model and making eternal inflation
impossible. Rather, finite volumes $\mathcal{V}$ are generated by
rare chance (i.e.~within the ensemble $E_{\mathcal{V}}$) through
the unmodified dynamics of the model, directly in the regime of eternal
inflation.

Below I compute the distribution $\rho(\mathcal{V})$ asymptotically
for very large $\mathcal{V}$ in models of slow-roll inflation (Sec.~\ref{sub:Asymptotics-of-rho}).
Specifically, I will compute the distribution $\rho(\mathcal{V};\phi_{0})$,
where $\phi_{0}$ is the (homogeneous) value of the inflaton field
in the initial region $S$. To implement the RV cutoff explicitly
for predicting the distribution of a cosmological parameter $Q$,
it is necessary to consider the joint finitely produced distribution
$\rho(\mathcal{V},\mathcal{V}_{Q_{R}};\phi_{0},Q_{0})$ for the reheating
volume $\mathcal{V}(R)$ and the portion $\mathcal{V}_{Q_{R}}$ of
the reheating volume in which $Q=Q_{R}$. (As before, $\phi_{0}$
and $Q_{0}$ are the values in the initial $H$-region.) If the distribution
$\rho(\mathcal{V},\mathcal{V}_{Q_{R}};\phi_{0},Q_{0})$ is found,
one can determine the mean volume $\left\langle \left.\mathcal{V}_{Q_{R}}\right|_{\mathcal{V}}\right\rangle $
while the total reheating volume $\mathcal{V}$ is held fixed,\begin{equation}
\left\langle \left.\mathcal{V}_{Q_{R}}\right|_{\mathcal{V}}\right\rangle =\frac{\int\rho(\mathcal{V},\mathcal{V}_{Q_{R}};\phi_{0},Q_{0})\mathcal{V}_{Q_{R}}d\mathcal{V}_{Q_{R}}}{\rho(\mathcal{V};\phi_{0},Q_{0})}.\end{equation}
 Then one computes the probability of finding the value of $Q$ within
the interval $\left[Q_{R},Q_{R}+dQ\right]$ at a random point in the
volume $\mathcal{V}$,\begin{equation}
p(Q=Q_{R};\mathcal{V})\equiv\frac{\left\langle \left.\mathcal{V}_{Q_{R}}\right|_{\mathcal{V}}\right\rangle }{\mathcal{V}}.\end{equation}
The RV cutoff \emph{defines} the probability distribution $p(Q)$
for an observable $Q$ as the limit of the distribution $p(Q;\mathcal{V})$
at large $\mathcal{V}$,\begin{equation}
p(Q)\equiv\lim_{\mathcal{V}\rightarrow\infty}\frac{\left\langle \left.\mathcal{V}_{Q_{R}}\right|_{\mathcal{V}}\right\rangle }{\mathcal{V}}.\label{eq:p Q def}\end{equation}
One expects that this limit is independent of the initial values $\phi_{0},Q_{0}$
because the large volume $\mathcal{V}$ is generated by regions that
spent a very long time in the self-reproduction regime and forgot
the initial conditions.

In Ref.~\cite{Winitzki:2008yb} I derived equations from which the
distributions $\rho(\mathcal{V},\mathcal{V}_{Q_{R}};\phi_{0},Q_{0})$
and $\rho(\mathcal{V};\phi_{0},Q_{0})$ can be in principle determined.
However, a direct computation of the limit $\mathcal{V}\rightarrow\infty$
(for instance, by a numerical method) will be cumbersome since the
relevant probabilities are exponentially small in that limit. One
of the main results of the present article is an analytic evaluation
of the limit $\mathcal{V}\rightarrow\infty$ and a derivation of a
more explicit formula, Eq.~(\ref{eq:pQR main ans}), for the distribution
$p(Q)$. The formula shows that the distribution $p(Q)$ can be computed
as a ground-state eigenfunction of a certain modified Fokker-Planck
equation. The explicit representation also proves that the limit~(\ref{eq:p Q def})
exists, is gauge-invariant, and is independent of the initial conditions
$\phi_{0}$ and $Q_{0}$. 

It was argued qualitatively in Ref.~\cite{Winitzki:2008yb} that
the RV measure does not suffer from the youngness paradox. In this
article I demonstrate the absence of the youngness paradox in the
RV measure by an explicit calculation. To this end, I will consider
a toy model where every $H$-region starts in the fluctuation-dominated
(or {}``self-reproduction'') regime with a constant expansion rate
$H_{0}$ and proceeds to reheating via two possible channels. The
first channel consists of a short period $\delta t_{1}$ of deterministic
slow-roll inflation, yielding $N_{1}$ $e$-folds until reheating;
the second channel has a different period $\delta t_{2}\neq\delta t_{1}$
of deterministic inflation, yielding $N_{2}$ $e$-folds. (For simplicity,
in this model one neglects fluctuations that may return the field
from the slow-roll regime to the self-reproduction regime, and thus
the time periods $\delta t_{1}$ and $\delta t_{2}$ are sharply defined.)
Thus there are two types of reheated regions corresponding to the
two possible slow-roll channels. The task is to compute the relative
volume-weighted probability $P(2)/P(1)$ of regions of these types
within the reheating surface. (Essentially the same model was considered,
e.g., in Refs.~\cite{Vilenkin:1995yd,Vilenkin:1998kr,Winitzki:2005ya,Linde:2007nm}.
See Fig.~\ref{cap:pot1} for a sketch of the potential $V(\phi)$in
this model.)

This toy model serves as a litmus test of measure prescriptions. The
{}``holographic'' or {}``worldline'' prescription yields $P(2)/P(1)$
equal to the probability ratio of exiting through the two channels
for a single comoving worldline. This probability ratio depends on
the initial conditions. Thus, the worldline measure is (by design)
blind to the volume growth during the slow-roll periods. On the other
hand, the volume-weighted prescriptions of Refs.~\cite{Vilenkin:1998kr,Linde:2007nm}
both yield \begin{equation}
\frac{P(2)}{P(1)}=\frac{\exp(3N_{2})}{\exp(3N_{1})},\label{eq:unbiased youngness}\end{equation}
rewarding the reheated $H$-regions that went through channel $j$
by the additional volume factor $\exp(3N_{j})$. This ratio is now
independent of the initial conditions. For comparison, an equal-time
cutoff gives\begin{equation}
\frac{P(2)}{P(1)}=\frac{\exp\left[3N_{2}-\left(3H_{\max}-\Gamma_{1}-\Gamma_{2}\right)\delta t_{2}\right]}{\exp\left[3N_{1}-\left(3H_{\max}-\Gamma_{1}-\Gamma_{2}\right)\delta t_{1}\right]}.\end{equation}
The overwhelming exponential dependence on $\delta t_{1}$ and $\delta t_{2}$
manifests the youngness paradox: Even a small difference $\delta t_{2}-\delta t_{1}$
in the duration of the slow-roll inflationary epoch leads to the exponential
bias towards the {}``younger'' universes. The bias persists regardless
of the choice of the time gauge~\cite{Winitzki:2005ya}, essentially
because the \emph{presence} of $\delta t_{1}$ and $\delta t_{2}$
in the ratio $P(2)/P(1)$ cannot be eliminated by using a different
time coordinate.%
\footnote{It should be noted that the youngness bias becomes very small, possibly
even negligible, if one uses the number $N$ of inflationary $e$-foldings
as the time variable rather than the proper time $t$. I am grateful
to A. Linde and A. Vilenkin for bringing this to my attention.%
} One expects that the RV measure will be free from this bias because
the RV prescription does not involve the time coordinate $t$ at all.
Below (Sec.~\ref{sub:Toy-model-of}) I will show that the ratio $P(2)/P(1)$
computed using the RV cutoff is indeed independent of the slow-roll
durations $\delta t_{1,2}$. The RV-regulated result {[}shown in Eq.~(\ref{eq:RV result toy})
below] depends only on the gauge-invariant quantities such as $N_{1}$
and $N_{2}$ and is, in general, different from Eq.~(\ref{eq:unbiased youngness}).
A calculation for an analogous landscape model was performed in Ref.~\cite{Winitzki:2008yb},
yielding a result qualitatively similar to Eq.~(\ref{eq:RV result toy}). 

These calculations confirm that the RV measure has the desirable properties
expected of a volume-based measure: coordinate invariance, independence
of initial conditions, and the absence of the youngness paradox. Thus
the RV measure is a promising solution to the long-standing problem
of obtaining probabilities in models of eternal inflation. Ultimately,
the viability of the RV measure proposal will depend on its performance
in various example cases. In the calculations available so far, it
is found that RV measure yields results that do not identically coincide
with the results of any other measure proposal. Hence, the RV measure
is not equivalent to earlier proposals and needs to be studied in
detail. 

As formulated here and in Ref.~\cite{Winitzki:2008yb}, the RV measure
prescription is directly applicable only to comparisons of reheating
volumes, or in general of \emph{terminal} states in the landscape
(such as the anti-de Sitter bubbles). The RV proposal needs to be
extended to predicting distributions of properties not directly related
to terminal states, such as the relative number of observations performed
in different nonterminal bubbles. Then it will be possible to investigate
whether the RV measure suffers from the {}``Boltzmann brain'' problem
or from other difficulties encountered by some previous measure proposals. 

An extension of the RV measure to landscape scenarios can be achieved
in several ways. For instance, one can consider the set of all possible
future evolutions of a single nonterminal bubble and define the ensemble
$E_{N}$ of evolutions yielding a finite total number $N$ of daughter
bubbles (of all types). One can also consider the ensemble $E_{N}^{\prime}$
of evolutions yielding a finite total number $N$ of \emph{observers}
in bubbles of all types. After computing the distribution of some
desired quantity by counting the observations made within the finite
set of $N$ bubbles (or observers), the cutoff parameter $N$ can
be increased to infinity. It remains to be seen whether the limit
distributions are different for differently defined ensembles, such
as $E_{N}$ and $E_{N}^{\prime}$, and if so, which definition is
more suitable. Future work will show whether some extension of the
RV measure can provide a satisfactory answer to the problem of predictions
in eternal inflation.

\section{Overview of the results}

In this section I describe the central results of this paper; in particular,
I develop simplified mathematical procedures for practical calculations
in the RV measure. For convenience of the reader, the results are
stated here without proof, while the somewhat lengthy derivations
are given in Sec.~\ref{sec:Derivations}.

\subsection{Preliminaries}

I consider a model of slow-roll inflation driven by an inflaton $\phi$
with the action \begin{equation}
\int\left[\frac{R}{16\pi G}+\frac{1}{2}\left(\partial_{\mu}\phi\right)^{2}-V(\phi)\right]\sqrt{-g}d^{4}x.\end{equation}
In the semiclassical stochastic approach to inflation,%
\footnote{See Refs.~\cite{Vilenkin:1983xq,Starobinsky:1986fx,Goncharov:1987ir}
for early works on the stochastic approach and Refs.~\cite{Linde:1993xx,Winitzki:2006rn}
for pedagogical reviews.%
} the semiclassical dynamics of the field $\phi$ averaged over an
$H$-region is regarded as a superposition of a deterministic slow
roll, \begin{equation}
\dot{\phi}=v(\phi)\equiv-\frac{V_{,\phi}(\phi)}{3H(\phi)}=-\frac{H_{,\phi}}{4\pi G},\label{eq:v def}\end{equation}
 and a random walk with root-mean-squared step size\begin{equation}
\sqrt{\left\langle \delta\phi\right\rangle ^{2}}=\frac{H(\phi)}{2\pi}\equiv\sqrt{\frac{2D(\phi)}{H(\phi)}},\quad D\equiv\frac{H^{3}}{8\pi^{2}},\label{eq:D def}\end{equation}
during time intervals $\delta t=H^{-1}$, where $H(\phi)$ is the
function defined by \begin{equation}
H(\phi)\equiv\sqrt{\frac{8\pi G}{3}V(\phi)}.\label{eq:H def}\end{equation}
A useful effective description of the evolution of the field at time
scales $\delta t\lesssim H^{-1}$ can be given as\begin{equation}
\phi(t+\delta t)=\phi(t)+v(\phi)\delta t+\xi(t)\sqrt{2D(\phi)\delta t},\label{eq:phi delta t}\end{equation}
where $\xi(t)$ is a normalized {}``white noise'' function, \begin{equation}
\left\langle \xi\right\rangle =0,\quad\left\langle \xi(t)\xi(t')\right\rangle =\delta(t-t'),\end{equation}
which is approximately statistically independent between different
$H$-regions. This stochastic process describes the evolution $\phi(t)$
and the accompanying cosmological expansion of space along a single
comoving worldline. For simplicity, we assume that inflation ends
in a given horizon-size region when $\phi=\phi_{*}$, where $\phi_{*}$
is a fixed value such that the relative change of $H$ during one
Hubble time $\delta t=H^{-1}$ becomes of order 1, i.e.\begin{equation}
\left|\frac{H_{,\phi}vH^{-1}}{H}\right|_{\phi=\phi_{*}}=\left|\frac{H_{,\phi}^{2}}{4\pi GH^{2}}\right|_{\phi=\phi_{*}}\sim1.\end{equation}
From the point of view of the stochastic approach, an inflationary
model is fully specified by the kinetic coefficients $D(\phi)$, $v(\phi)$,
$H(\phi)$. These coefficients are found from Eqs.~(\ref{eq:v def})--(\ref{eq:H def})
in models of canonical slow-roll inflation and by suitable analogues
in other models.

Dynamics of any fluctuating cosmological parameter $Q$ is described
in a similar way. One assumes that the value of $Q$ is homogeneous
in $H$-regions. The evolution of $Q$ is described by an effective
Langevin equation,\begin{equation}
Q(t+\delta t)=Q(t)+v_{Q}(\phi,Q)\delta t+\xi_{Q}(t)\sqrt{2D_{Q}(\phi,Q)\delta t},\end{equation}
where the kinetic coefficients $D_{Q}$ and $v_{Q}$ can be computed,
similarly to $D$ and $v$, from first principles. For simplicity
we assume that the {}``noise variable'' $\xi_{Q}$ is independent
of the {}``noise'' $\xi$ used in Eq.~(\ref{eq:phi delta t}).
A correlated set of noise variables can be considered as well (see
e.g.~Ref.~\cite{Tolley:2008na}).

\subsection{Probability of finite inflation}

Let us consider an initial $H$-region $S$ where the inflaton field
$\phi$ as well as the parameter $Q$ are homogeneous and have values
$\phi=\phi_{0}$ and $Q=Q_{0}$. For convenience we assume that reheating
starts when $\phi=\phi_{*}$ and the Planck energy scales are reached
at $\phi=\phi_{\text{Pl}}$ independently of the value of $Q$. (If
necessary, the field variables $\phi,Q$ can be redefined to achieve
this.)

Although eternal inflation to the future of $S$ is almost always
the case, it is possible that reheating is reached at a finite time
everywhere to the future of $S$, due to a rare fluctuation. In that
event, the total reheating volume $\mathcal{V}$ to the future of
$S$ is finite. The (small) probability of that event, denoted by
\begin{equation}
\text{Prob}\left(\mathcal{V}<\infty|\phi_{0},Q_{0}\right)\equiv\bar{X}(\phi_{0},Q_{0}),\end{equation}
 can be found as the solution of the following nonlinear equation,
\begin{align}
\frac{D}{H}\bar{X}_{,\phi\phi}+\frac{D_{Q}}{H}\bar{X}_{,QQ}+\frac{v}{H}\bar{X}_{,\phi}+\frac{v_{Q}}{H}\bar{X}_{,Q}+3\bar{X}\ln\bar{X} & =0,\label{eq:X equ}\\
\bar{X}(\phi_{\text{Pl}},Q)=1,\quad\bar{X}(\phi_{*},Q) & =1,\end{align}
where for brevity we dropped the subscript 0 in $\phi_{0}$ and $Q_{0}$.
This basic equation, first derived in Ref.~\cite{Winitzki:2001np},
is of reaction-diffusion type and can be viewed as a nonlinear modification
of the Fokker-Planck equations used previously in the literature on
the stochastic approach to inflation.

While $\bar{X}(\phi,Q)\equiv1$ is always a solution of Eq.~(\ref{eq:X equ}),
it is not the correct one for the case of eternal inflation. A nontrivial
solution, $\bar{X}(\phi,Q)\not\equiv1$, exists and has small values
$\bar{X}(\phi,Q)\ll1$ for $\phi,Q$ away from the thermalization
boundary. If the coefficients $D/H$ and $v/H$ happen to be $Q$-independent,
the solution of Eq.~(\ref{eq:X equ}) will be also independent of
$Q$, i.e.~$\bar{X}(\phi,Q)=\bar{X}(\phi)$, and thus determined
by a simpler equation obtained from Eq.~(\ref{eq:X equ}) by omitting
derivatives with respect to $Q$,\begin{equation}
\frac{D}{H}\bar{X}_{,\phi\phi}+\frac{v}{H}\bar{X}_{,\phi}+3\bar{X}\ln\bar{X}=0.\label{eq:X equ phi}\end{equation}
It is easy to see that Eqs.~(\ref{eq:X equ}) and (\ref{eq:X equ phi})
are manifestly gauge-invariant. Indeed, a change of time variable
according to Eq.~(\ref{eq:general time gauge}) results in dividing
the coefficients $D,D_{Q},v,v_{Q},H$ by the function $T(\phi)$~\cite{Winitzki:1995pg},
which leaves Eqs.~(\ref{eq:X equ}) and (\ref{eq:X equ phi}) unmodified.

Some approximate solutions of Eq.~(\ref{eq:X equ phi}) were given
in Ref.~\cite{Winitzki:2001np}, where it was shown that $\bar{X}(\phi)$
is typically exponentially small for $\phi$ in the inflationary regime.
While small, $\bar{X}(\phi)$ is never zero; hence, there is a well-defined
statistical ensemble of initial $H$-regions that have a finite total
reheating volume in the future. The construction of the RV measure
relies on this fact.

\subsection{Finitely produced volume}

In a scenario where eternal inflation is possible, we now consider
the probability density $\rho(\mathcal{V};\phi_{0})$ of having a
\emph{finite} total reheating volume $\mathcal{V}$ to the comoving
future of an initial $H$-region with homogeneous value $\phi=\phi_{0}$
(focusing attention at first on the case of inflation driven by a
single scalar field). The distribution $\rho(\mathcal{V};\phi_{0})$
is normalized to the overall probability $\bar{X}(\phi_{0})$ of having
a finite total reheating volume, \begin{equation}
\int_{0}^{\infty}\rho(\mathcal{V};\phi_{0})d\mathcal{V}=\bar{X}(\phi_{0}).\end{equation}
The distribution $\rho(\mathcal{V};\phi_{0})$ can be calculated by
first determining the generating function $g(z;\phi_{0})$, which
is defined by\begin{equation}
g(z;\phi_{0})\equiv\left\langle e^{-z\mathcal{V}}\right\rangle _{\mathcal{V}<\infty}\equiv\int_{0}^{\infty}e^{-z\mathcal{V}}\rho(\mathcal{V};\phi_{0})d\mathcal{V}.\label{eq:g def 0}\end{equation}
This generating function is a solution of the nonlinear Fokker-Planck
equation, \begin{equation}
\hat{L}g+3g\ln g=0,\label{eq:g equ 0}\end{equation}
where the differential operator $\hat{L}$ is defined by\begin{equation}
\hat{L}\equiv\frac{D}{H}\partial_{\phi}\partial_{\phi}+\frac{v}{H}\partial_{\phi}.\end{equation}
In the case of several fields, say $\phi$and $Q$, one needs to use
the corresponding Fokker-Planck operator such as\begin{equation}
\hat{L}=\frac{D_{\phi\phi}}{H}\partial_{\phi}\partial_{\phi}+\frac{D_{QQ}}{H}\partial_{Q}\partial_{Q}+\frac{v_{\phi}}{H}\partial_{\phi}+\frac{v_{Q}}{H}\partial_{Q}.\label{eq:L FP phi Q}\end{equation}
The boundary conditions for Eq.~(\ref{eq:g equ 0}) are\begin{align}
g(z;\phi,Q) & =1\:\mbox{\text{for} }\left\{ \phi,Q\right\} \in\text{Planck boundary},\\
g(z;\phi,Q) & =\left.e^{-zH^{-3}(\phi,Q)}\right|_{\left\{ \phi,Q\right\} \in\text{reheating boundary}}.\end{align}
Note that the parameter $z$ enters the boundary conditions but is
not explicitly involved in Eq.~(\ref{eq:g equ 0}). Also, the operator
$\hat{L}$ and Eq.~(\ref{eq:g equ 0}) are manifestly gauge-invariant
with respect to redefinitions of the form~(\ref{eq:general time gauge}).

The generating function $g$ plays a central role in the calculations
of the RV cutoff. It will be shown below that the solution $g(z;\phi,Q)$
of Eq.~(\ref{eq:g equ 0}) needs to be obtained only at an appropriately
determined \emph{negative} value of $z$. This solution can be obtained
by a numerical method or through an analytic approximation if available.

\subsection{Asymptotics of $\rho(\mathcal{V};\phi_{0})$\label{sub:Asymptotics-of-rho}}

The finitely produced distribution $\rho(\mathcal{V};\phi)$ can be
found through the inverse Laplace transform of the function $g(z;\phi)$,\begin{equation}
\rho(\mathcal{V};\phi)=\frac{1}{2\pi\text{i}}\int_{-\text{i}\infty}^{\text{i}\infty}\negmedspace dz\, e^{z\mathcal{V}}g(z;\phi),\label{eq:rho integral 0}\end{equation}
where the integration contour in the complex $z$ plane can be chosen
along the imaginary axis. The asymptotic behavior of $\rho(\mathcal{V};\phi)$
at large $\mathcal{V}$ is determined by the right-most singularity
of $g(z;\phi)$ in the complex $z$ plane. It turns out that the function
$g(z;\phi)$ always has a singularity at a real, nonpositive $z=z_{*}$
of the type\begin{equation}
g(z;\phi)=g(z_{*};\phi)+\sigma(\phi)\sqrt{z-z_{*}}+O(z-z_{*}),\label{eq:g sing}\end{equation}
where $z_{*}$ and $\sigma(\phi)$ are determined as follows. One
considers the ($z$-dependent) linear operator\begin{equation}
\hat{\tilde{L}}\equiv\hat{L}+3(\ln g(z;\phi)+1),\end{equation}
where $\hat{L}$ is the Fokker-Planck operator described above. For
$z>0$ this operator is invertible in the space of functions $f(\phi)$
satisfying zero boundary conditions. The value of $z_{*}$ turns out
to be the algebraically largest real number (in any case, $z_{*}\leq0$)
such that there exists an eigenfunction $\sigma(\phi)$ of $\hat{\tilde{L}}$
with zero eigenvalue and zero boundary conditions,\begin{equation}
\hat{\tilde{L}}\sigma(\phi)=0,\quad\sigma(\phi_{*})=\sigma(\phi_{\text{Pl}})=0.\end{equation}
The specific normalization of the eigenfunction $\sigma(\phi)$ can
be derived analytically but is unimportant for the present calculations.

The singularity type shown in Eq.~(\ref{eq:g sing}) determines the
leading asymptotic of $\rho(\mathcal{V};\phi)$ at $\mathcal{V}\rightarrow\infty$:
\begin{equation}
\rho(\mathcal{V};\phi)\approx\frac{1}{2\sqrt{\pi}}\sigma(\phi)\mathcal{V}^{-3/2}e^{z_{*}\mathcal{V}}.\label{eq:rho asympt}\end{equation}
The explicit form~(\ref{eq:rho asympt}) allows one to investigate
the moments of the distribution $\rho(\mathcal{V};\phi)$. It is clear
that all the moments are finite as long as $z_{*}<0$. However, if
$z_{*}=0$ all the moments diverge, namely for $n\geq1$ we have\begin{equation}
\left\langle \mathcal{V}^{n}\right\rangle =\int_{0}^{\infty}\rho(\mathcal{V};\phi)\mathcal{V}^{n}d\mathcal{V}\propto\int_{0}^{\infty}\mathcal{V}^{n-3/2}d\mathcal{V}=\infty.\end{equation}
The case $z_{*}=0$ corresponds to the {}``transition point'' analyzed
in Ref.~\cite{Creminelli:2008es}, corresponding to $\Omega=1$ in
their notation. This is the borderline case between the presence and
the absence of eternal inflation. The fact that $z_{*}=0$ in the
borderline case can be seen directly by noting that the Fokker-Planck
operator $\hat{L}+3$ has in that case a zero eigenvalue, meaning
that the 3-volume of equal-time surfaces does not expand with time
(reheating of some regions is perfectly compensated by inflationary
expansion of other regions). In that case, the only solution $g(z=0;\phi)=\bar{X}(\phi)$
of Eq.~(\ref{eq:X equ phi}) is $\bar{X}\equiv1$ because there are
no eternally inflating comoving geodesics. Hence $\ln g(z=0,\phi)=0$,
and so the operator $\hat{\tilde{L}}$ is simply $\hat{\tilde{L}}=\hat{L}+3$.
It follows that the operator $\hat{\tilde{L}}$ also has a zero eigenvalue
at $z=0$, and thus $z=z_{*}=0$ is the dominant singularity of $g(z;\phi)$.
This argument reproduces and generalizes the results obtained in Ref.~\cite{Creminelli:2008es}
where direct calculations of various moments of $\rho(\mathcal{V};\phi)$
were performed for the case of the absence of eternal inflation.

We note that the only necessary ingredients in the computation of
$\sigma(\phi)$ is the knowledge of the singularity point $z_{*}$
and the corresponding function $g(z_{*};\phi)$, which is a solution
of the nonlinear reaction-diffusion equation~(\ref{eq:g equ 0}).
Determining $z_{*}$ and $g(z_{*};\phi)$ in a given inflationary
model does not require extensive numerical simulations.

\subsection{Distribution of a fluctuating field}

Above we denoted by $Q$ a cosmological parameter that fluctuates
during inflation but is in principle observable after reheating. One
of the main questions to be answered using a multiverse measure is
to derive the probability distribution $p(Q)$ for the values of $Q$
observed in a {}``typical'' place in the multiverse. I will now
present a formula for the distribution $p(Q)$ in the RV cutoff. This
formula is significantly more explicit and lends itself more easily
to practical calculations than the expressions first shown in Ref.~\cite{Winitzki:2008yb}.

As in the previous section, we assume that the dynamics of the inflaton
field $\phi$ and the parameter $Q$ is described by a suitable Fokker-Planck
operator $\hat{L}$, e.g.~of the form~(\ref{eq:L FP phi Q}), and
that reheating occurs at $\phi=\phi_{*}$ independently of the value
of $Q$. We then consider Eq.~(\ref{eq:g equ 0}) for the function
$g(z;\phi,Q)$ and the operator $\hat{\tilde{L}}\equiv\hat{L}+3(\ln g+1)$;
we need to determine the value $z_{*}$ at which $g(z;\phi,Q)$ has
a singularity. The operator $\hat{\tilde{L}}$ has an eigenfunction
with zero eigenvalue for this value of $z$. This eigenfunction $f_{0}(z_{*};\phi,Q)$
needs to be determined with zero boundary conditions (at reheating
and Planck boundaries). Then the RV-regulated distribution of $Q$
at reheating is \begin{equation}
p(Q_{R})=\mbox{const}\left[\frac{\partial f_{0}(z_{*};\phi,Q)}{\partial\phi}\frac{D_{\phi\phi}e^{-z_{*}H^{-3}}}{H^{4}}\right]_{\phi=\phi_{*},Q=Q_{R}},\label{eq:pQR main ans}\end{equation}
where the normalization constant needs to be chosen such that $\int p(Q_{R})dQ_{R}=1$.
The derivation of this result occupies Sec.~\ref{sub:FPRV-distribution-of-Q}. 

We note that $f_{0}$ is the eigenfunction $f_{0}$ of a gauge-invariant
operator, and that the result in Eq.~(\ref{eq:pQR main ans}) depends
on the kinetic coefficients only through the gauge-independent ratio
$D/H$ times the volume factor $H^{-3}$. The distribution $p(Q_{R})$
is independent of the initial conditions, which is due to a specific
asymptotic behavior of the finitely produced volume distributions,
as shown in Sec.~\ref{sub:FPRV-distribution-of-Q}.

\subsection{Toy model of inflation\label{sub:Toy-model-of}}

\begin{figure}
\begin{centering}\psfrag{phi}{$\phi$}\psfrag{V}{$V$}

\psfrag{ps2}{$\phi_{*}^{(1)}$}\psfrag{ps1}{$\phi_{*}^{(2)}$}

\psfrag{p2}{$\phi_1$}\psfrag{p1}{$\phi_2$}\includegraphics[width=3.5in]{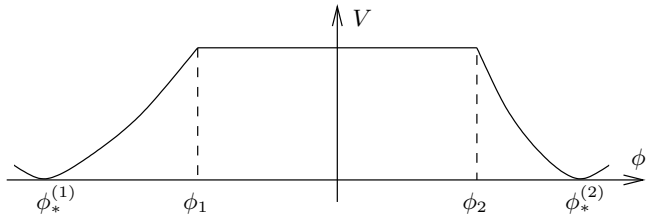}\par\end{centering}

\caption{A model potential with a flat self-reproduction regime $\phi_{1}<\phi<\phi_{2}$
and deterministic slow-roll regimes $\phi_{*}^{(1)}<\phi<\phi_{1}$
and $\phi_{2}<\phi<\phi_{*}^{(2)}$ producing $N_{1}$ and $N_{2}$
inflationary $e$-folds respectively. In the interval $\phi_{1}<\phi<\phi_{2}$
the potential $V(\phi)$ is assumed to be constant, $V(\phi)=V_{0}$.
\label{cap:pot1}}
\end{figure}

We now apply the RV cutoff to the toy model described at the end of
Sec.~\ref{sec:Introduction-and-motivation}. We consider a model
of inflation driven by a scalar field with a potential shown in Fig.~\ref{cap:pot1}.
For the purposes of the present argument, we may assume that there
is exactly zero {}``diffusion'' in the deterministic regimes $\phi_{*}^{(1)}<\phi<\phi_{1}$
and $\phi_{2}<\phi<\phi_{*}^{(2)}$, while the range $\phi_{1}<\phi<\phi_{2}$
is sufficiently wide to allow for eternal self-reproduction. Thus
there are two slow-roll channels that produce respectively $N_{1}$
and $N_{2}$ $e$-folds of slow-roll inflation after exiting the self-reproduction
regime. Since the self-reproduction range generates arbitrarily large
volumes of space that enter both the slow-roll channels, the total
reheating volume going through each channel is infinite. We apply
the RV cutoff to the problem of computing the regularized ratio of
the reheating volumes in regions of types 1 and 2. 

In this toy model it is possible to obtain the results of the RV cutoff
using analytic approximations. The required calculations are somewhat
lengthy and can be found in Sec.~\ref{sub:A-model-of-slow-roll}.
The result for a generic case where one of the slow-roll channels
has many more $e$-folds than the other (say, $N_{2}\gg N_{1}$) can
be written as\begin{equation}
\frac{P(2)}{P(1)}\approx O(1)\frac{H^{-3}(\phi_{*}^{(2)})}{H^{-3}(\phi_{*}^{(1)})}\frac{\exp\left[3N_{2}\right]}{\exp\left[3N_{1}\right]}\exp\left[3N_{12}\right],\label{eq:RV result toy}\end{equation}
where we have defined \begin{equation}
N_{12}\equiv\frac{\pi^{2}}{\sqrt{2}H_{0}^{2}}\left(\phi_{2}-\phi_{1}\right)^{2},\quad H_{0}^{2}\equiv\frac{8\pi G}{3}V_{0}.\end{equation}
The pre-exponential factor $O(1)$ can be computed numerically, as
outlined in Sec.~\ref{sub:A-model-of-slow-roll}. 

We note that the ratio~(\ref{eq:RV result toy}) is gauge-invariant
and does not involve any spacetime coordinates. This result can be
interpreted as the ratio of volumes $e^{3N_{1}}$ and $e^{3N_{2}}$
gained during the slow-roll regime in the two channels multiplied
by a correction factor $e^{3N_{12}}$. The dimensionless number $N_{12}$
can be suggestively interpreted (up to the factor $\sqrt{2}$) as
the mean number of {}``steps'' of size $\delta\phi\sim\frac{1}{2\pi}H_{0}$
required for a random walk to reach the boundary of the flat region
$\left[\phi_{1},\phi_{2}\right]$ starting from the middle point $\phi_{0}\equiv\frac{1}{2}\left(\phi_{1}+\phi_{2}\right)$.
Since each of the {}``steps'' of the random walk takes a Hubble
time $H_{0}^{-1}$ and corresponds to one $e$-folding of inflation,
the volume factor gained during such a traversal will be $e^{3N_{12}}$.
Note that the correction factor increases the probability of channel
2 that was \emph{already} the dominant one due to the larger volume
factor $e^{3N_{2}}\gg e^{3N_{1}}$. Depending on the model, this factor
may be a significant modification of the ratio~(\ref{eq:unbiased youngness})
obtained in previously used volume-based measures.

\section{Derivations\label{sec:Derivations}}

\subsection{Positive solutions of nonlinear equations}

It is not easy to demonstrate directly the existence of nontrivial
solutions of reaction-diffusion equations such as Eq.~(\ref{eq:X equ}).
However, there is a connection between solutions of such nonlinear
equations and solutions of the linearized equations. Rigorous results
are available in the mathematical literature on nonlinear functional
analysis and bifurcation theory.

Heuristically, consider a solution of Eq.~(\ref{eq:X equ phi}) that
is approximately $\bar{X}(\phi)\approx1$. The equation can be linearized
in the neighborhood of $\bar{X}\approx1$ as $\bar{X}=1-\chi(\phi)$
and yields the Fokker-Planck (FP) equation\begin{equation}
\left[\hat{L}+3\right]\chi=0,\quad\hat{L}\equiv\frac{D}{H}\partial_{\phi\phi}+\frac{v}{H}\partial_{\phi}.\label{eq:chi equ}\end{equation}
The FP operator $\hat{L}+3$ is adjoint to the operator\begin{equation}
\left[\hat{L}^{\dagger}+3\right]P\equiv\partial_{\phi\phi}\left(\frac{D}{H}P\right)-\partial_{\phi}(\frac{v}{H}P)+3P,\end{equation}
which enters the FP equation for the 3-volume distribution $P(\phi,t)$
in the $e$-folding time parameterization. If eternal inflation is
allowed in a given model, the operator $\hat{L}^{\dagger}+3$ has
a positive eigenvalue. The largest eigenvalue of that operator is
zero in the borderline case when eternal inflation is just about to
set in. The spectrum of the operator $\hat{L}+3$ is the same as that
of the adjoint operator $\hat{L}^{\dagger}+3$. Hence, in the borderline
case the largest eigenvalue of the operator $\hat{L}+3$ will be zero,
and there will exist a nontrivial, everywhere nonnegative solution
$\chi$ of Eq.~(\ref{eq:chi equ}). Thus, heuristically one can expect
that a nontrivial solution $\bar{X}(\phi)\not\equiv1$ will exist
away from the borderline case, i.e.~when the operator $\hat{L}+3$
has a positive eigenvalue.

Following the approach of Ref.~\cite{Creminelli:2008es}, one can
imagine a family of inflationary models parameterized by a label $\Omega$,
such that eternal inflation is allowed when $\Omega<1$. Then Eq.~(\ref{eq:X equ phi})
will have only the trivial solution, $\bar{X}(\phi)\equiv1$, for
$\Omega\geq1$. The case $\Omega=1$ where eternal inflation is on
the borderline of existence is the \emph{bifurcation point} for the
solutions of Eq.~(\ref{eq:X equ phi}). At the bifurcation point,
a nontrivial solution $\bar{X}(\phi)\not\equiv1$ appears, branching
off from the trivial solution. A rigorous theory of bifurcation can
be developed using methods of nonlinear functional analysis (see e.g.~chapter
9 of the book~\cite{Stakgold:1979:GF}). In particular, it can be
shown that a nontrivial solution of a nonlinear equation, such as
Eq.~(\ref{eq:X equ phi}), exists if and only if the dominant eigenvalue
of the linearized operator $\hat{L}+3$ with zero boundary conditions
is positive. 

There remains a technical difference between the eigenvalue problem
for the operator $\hat{L}+3$ with zero boundary conditions and with
the {}``no-diffusion'' boundary conditions normally used in the
stochastic approach,\begin{equation}
\left.\frac{\partial}{\partial\phi}\right|_{\phi_{*}}\left[D(\phi)P(\phi)\right]=0.\label{eq:no diff bc}\end{equation}
It was demonstrated in Ref.~\cite{Winitzki:2001np} that the eigenvalue
of $\hat{L}+3$ with the boundary conditions~(\ref{eq:no diff bc})
is positive if a nontrivial solution of Eq.~(\ref{eq:X equ phi})
exists. In principle, the eigenvalue of $\hat{L}+3$ with zero boundary
conditions is not the same as the eigenvalue of the same operator
with the boundary conditions~(\ref{eq:no diff bc}). One can have
a borderline case when one of these two eigenvalues is positive while
the other is negative. In this case, the two criteria for the presence
of eternal inflation (based on the positivity of the two different
eigenvalues) will disagree. However, the alternative boundary conditions
are imposed at reheating, i.e.~in the regime of very small fluctuations
where the value of the eigenfunction $P(\phi)$ is exponentially small
compared with its values in the fluctuation-dominated range of $\phi$.
Hence, the difference between the two eigenvalues is always exponentially
small (it is suppressed at least by the factor $e^{-3N}$, where $N$
is the number of $e$-folds in the deterministic slow-roll regime
before reheating). Therefore, we may interpret the discrepancy as
a limitation inherent in the stochastic approach to inflation. In
other words, one cannot use the stochastic approach to establish the
presence of eternal inflation more precisely than with the accuracy
$e^{-3N}$. Barring an extremely fine-tuned borderline case, this
accuracy is perfectly adequate for establishing the presence or absence
of eternal inflation.

The main nonlinear equation in the calculations of the RV cutoff is
Eq.~(\ref{eq:g equ 0}) for the generating function $g(z;\phi)$.
That equation differs from Eq.~(\ref{eq:X equ phi}) mainly by the
presence of the parameter $z$ in the boundary conditions. Therefore,
solutions of Eq.~(\ref{eq:g equ 0}) may exist for some values of
$z$ but not for other values. Note that $g(z=0;\phi)=\bar{X}(\phi)$;
hence, nontrivial solutions $g(z;\phi)$ exist for $z=0$ under the
same conditions as nontrivial solutions $\bar{X}(\phi)\not\equiv1$
of Eq.~(\ref{eq:X equ phi}). While it is certain that solutions
$g(z;\phi)$ exist for $z\geq0$, there may be values $z<0$ for which
no real-valued solutions $g(z;\phi)$ exist at all. However, the calculations
in the RV cutoff require only to compute $g(z_{*};\phi)$ for a certain
value $z_{*}<0$, which is the algebraically largest value $z$ where
$g(z;\phi)$ has a singularity in the $z$ plane. The structure of
that singularity will be investigated in detail below, and it will
be shown that $g(z_{*};\phi)$ is finite while $\partial g/\partial z\propto(z-z_{*})^{-1/2}$
diverges at $z=z_{*}$. Hence, the solution $g(z;\phi)$ remains well-defined
at least for all real $z$ in the interval $\left[z_{*},+\infty\right]$.
It follows that $g(z;\phi)$ may be obtained e.g.~by a numerical
solution of a well-conditioned problem with $z=z_{*}+\varepsilon$,
where $\varepsilon>0$ is a small real constant.

\subsection{Nonlinear Fokker-Planck equations\label{sub:Distribution-of-finitely-produced}}

In this section I derive Eq.~(\ref{eq:g equ 0}), closely following
the derivation of Eq.~(\ref{eq:X equ}) in Ref.~\cite{Winitzki:2001np}. 

We begin by considering the case when inflation is driven by a single
scalar field $\phi$, such that reheating is reached at $\phi=\phi_{*}$.
Let $\rho(\mathcal{V};\phi_{0})$ be the probability density of obtaining
the finite reheated volume $\mathcal{V}$. We will derive an equation
for a generating function of the distribution of volume, rather than
an equation directly for $\rho(\mathcal{V};\phi_{0})$. Since the
volume $\mathcal{V}$ is by definition nonnegative, it is convenient
to define a generating function $g(z;\phi_{0})$ through the expectation
value of the expression $\exp(-z\mathcal{V})$, where $z>0$ is the
formal parameter of the generating function, \begin{equation}
g(z;\phi_{0})\equiv\left\langle e^{-z\mathcal{V}}\right\rangle _{\mathcal{V}<\infty}\equiv\int_{0}^{\infty}e^{-z\mathcal{V}}\rho(\mathcal{V};\phi_{0})d\mathcal{V}.\label{eq:g def}\end{equation}
Note that for any $z$ such that $\mbox{Re}\, z\geq0$ the integral
in Eq.~(\ref{eq:g def}) converges, and the events with $\mathcal{V}=+\infty$
are automatically excluded from consideration. However, we use the
subscript {}``$\mathcal{V}<\infty$'' to indicate explicitly that
the statistical average is performed over a subset of all events.
The distribution $\rho(\mathcal{V};\phi_{0})$ is not normalized to
unity; instead, the normalization is given by Eq.~(\ref{eq:rho V normalization}).

The parameter $z$ has the dimension of inverse 3-volume. Physically,
this is the 3-volume measured along the reheating surface and hence
is defined in a gauge-invariant manner. If desired for technical reasons,
the variable $z$ can be made dimensionless by a constant rescaling.

The generating function $g(z;\phi)$ has the following multiplicative
property: For two statistically independent regions that have initial
values $\phi=\phi_{1}$ and $\phi=\phi_{2}$ respectively, the sum
of the (finitely produced) reheating volumes $\mathcal{V}_{1}+\mathcal{V}_{2}$
is distributed with the generating function \begin{equation}
\left\langle e^{-z(\mathcal{V}_{1}+\mathcal{V}_{2})}\right\rangle =\left\langle e^{-z\mathcal{V}_{1}}\right\rangle \left\langle e^{-z\mathcal{V}_{2}}\right\rangle =g(z;\phi_{1})g(z;\phi_{2}).\label{eq:g multiplicative}\end{equation}

We now consider an $H$-region at some time $t$, having an arbitrary
value $\phi(t)$ not yet in the reheating regime. Suppose that the
finitely produced volume distribution for this $H$-region has the
generating function $g(z;\phi)$. After time $\delta t$ the initial
$H$-region grows to $N\equiv e^{3H\delta t}$ statistically independent,
{}``daughter'' $H$-regions. The value of $\phi$ in the $k$-th
daughter region ($k=1,...,N$) is found from Eq.~(\ref{eq:phi delta t}),
\begin{equation}
\phi_{k}=\phi+v(\phi)\delta t+\xi_{k}\sqrt{2D(\phi)\delta t},\end{equation}
 where the {}``noise'' variables $\xi_{k}$ ($k=1,...,N$) are statistically
independent because they describe the fluctuations of $\phi$ in causally
disconnected $H$-regions. The finitely produced volume distribution
for the $k$-th daughter region has the generating function $g(z;\phi_{k})$.
The combined reheating volume of the $N$ daughter regions must be
distributed with the same generating function as reheating volume
of the original $H$-region. Hence, by the multiplicative property
we obtain\begin{equation}
g(z;\phi)=\prod_{k=1}^{N}g(z;\phi_{k}).\end{equation}
 We can average both sides of this equation over the noise variables
$\xi_{k}$ to get\begin{equation}
g(z;\phi)=\left\langle \prod_{k=1}^{N}g(z;\phi_{k})\right\rangle _{\xi_{1},...,\xi_{N}}.\end{equation}
Since all the $\xi_{k}$ are independent, the average splits into
a product of $N$ identical factors, \begin{equation}
g(z;\phi)=\left[\left\langle g(z;\phi+v(\phi)\delta t+\sqrt{2D(\phi)}\xi)\right\rangle _{\xi}\right]^{N}.\label{eq:g equ pre}\end{equation}
The derivation now proceeds as in Ref.~\cite{Winitzki:2001np}. We
first compute, to first order in $\delta t$,\begin{equation}
\left\langle g(z;\phi+v(\phi)\delta t+\sqrt{2D(\phi)}\xi)\right\rangle _{\xi}=g+\left(vg_{,\phi}+Dg_{,\phi\phi}\right)\delta t.\end{equation}
 Substituting $N=e^{3H\delta t}$ and taking the logarithmic derivative
of both sides of Eq.~(\ref{eq:g equ pre}) with respect to $\delta t$
at $\delta t=0$, we then obtain\begin{align}
0 & =\frac{\partial}{\partial\delta t}\ln g(z;\phi)\nonumber \\
 & =3H\ln g+\frac{vg_{,\phi}+Dg_{,\phi\phi}}{g}.\end{align}
The equation for $g(z;\phi)$ follows,\begin{equation}
Dg_{,\phi\phi}+vg_{,\phi}+3Hg\ln g=0.\label{eq:g equ}\end{equation}
This is formally the same as Eq.~(\ref{eq:X equ}). However, the
boundary conditions for Eq.~(\ref{eq:g equ}) are different. The
condition at the end-of-inflation boundary $\phi=\phi_{*}$ is\begin{equation}
g(z;\phi_{*})=e^{-zH^{-3}(\phi_{*})}\label{eq:bc reheating}\end{equation}
because an $H$-region starting with $\phi=\phi_{*}$ immediately
reheats and produces the reheating volume $H^{-3}(\phi_{*})$. The
condition at Planck boundary $\phi_{\text{Pl}}$ (if present), or
other boundary where the effective field theory breaks down, is {}``absorbing,''
i.e.~regions that reach $\phi=\phi_{\text{Pl}}$ do not generate
any reheating volume: \begin{equation}
g(z;\phi_{\text{Pl}})=1.\label{eq:bc Planck}\end{equation}
The variable $z$ enters Eq.~(\ref{eq:g equ}) as a parameter and
only through the boundary conditions. At $z=0$ the solution is $g(0;\phi)=\bar{X}(\phi)$.

A fully analogous derivation can be given for the generating function
$g(z;\phi_{0},Q_{0})$ in the case when additional fluctuating fields,
denoted by $Q$, are present. The generating function $g(z;\phi_{0},Q_{0})$
is defined by\begin{equation}
g(z;\phi_{0},Q_{0})=\int_{0}^{\infty}e^{-z\mathcal{V}}\rho(\mathcal{V};\phi_{0},Q_{0})d\mathcal{V},\end{equation}
where $\rho(\mathcal{V};\phi_{0},Q_{0})$ is the probability density
for achieving a total reheating volume $\mathcal{V}$ in the future
of an $H$-region with initial values $\phi_{0},Q_{0}$ of the fields.
In the general case, the fluctuations of the fields $\phi,Q$ can
be described by the Langevin equations\begin{align}
\phi(t+\delta t) & =\phi(t)+v_{\phi}\delta t+\xi_{\phi}\sqrt{2D_{\phi\phi}\delta t}+\xi_{Q}\sqrt{2D_{\phi Q}\delta t},\\
Q(t+\delta t) & =Q(t)+v_{Q}\delta t+\xi_{\phi}\sqrt{2D_{\phi Q}\delta t}+\xi_{Q}\sqrt{2D_{QQ}\delta t},\end{align}
 where the {}``diffusion'' coefficients $D_{\phi\phi}$, $D_{\phi Q}$,
and $D_{QQ}$ have been introduced, as well as the {}``slow roll''
velocities $v_{\phi}$ and $v_{Q}$ and the {}``noise'' variables
$\xi_{\phi}$ and $\xi_{Q}$. The resulting equation for $g(z;\phi_{0},Q_{0})$
is (dropping the subscript 0)\begin{equation}
\hat{L}g+3g\ln g=0,\label{eq:g equ gen 1}\end{equation}
where the differential operator $\hat{L}$ is defined by \begin{equation}
\hat{L}\equiv\frac{D_{\phi\phi}}{H}\partial_{\phi}\partial_{\phi}+\frac{2D_{\phi Q}}{H}\partial_{\phi}\partial_{Q}+\frac{D_{QQ}}{H}\partial_{Q}\partial_{Q}+\frac{v_{\phi}}{H}\partial_{\phi}+\frac{v_{Q}}{H}\partial_{Q}.\label{eq:FP operator L for Q}\end{equation}
The ratios $D_{\phi\phi}/H$, etc., are manifestly gauge-invariant
with respect to time parameter changes of the form~(\ref{eq:general time gauge}).

Performing a redefinition of the fields if needed, one may assume
that reheating is reached when $\phi=\phi_{*}$ independently of the
value of $Q$. Then the boundary conditions for Eq.~(\ref{eq:g equ gen 1})
at the reheating boundary can be written as \begin{equation}
g(z;\phi_{*},Q)=e^{-zH^{-3}(\phi_{*},Q)}.\label{eq:bc reheating phi Q}\end{equation}
The Planck boundary still has the boundary condition $g(z;\phi_{\text{Pl}})=1$.

\subsection{Singularities of $g(z)$\label{sub:Singularities-of-g}}

For simplicity we now focus attention on the case of single-field
inflation; the generating function $g(z;\phi)$ then depends on the
initial value of the inflaton field $\phi$. The corresponding analysis
for multiple fields is carried out as a straightforward generalization.

By definition, $g(z;\phi)$ is an integral of a probability distribution
$\rho(\mathcal{V};\phi)$ times $e^{-z\mathcal{V}}$. It follows that
$g(z;\phi)$ is analytic in $z$ and has no singularities for $\mbox{Re}\, z>0$.
Then the probability distribution $\rho(\mathcal{V};\phi)$ can be
recovered from the generating function $g(z;\phi)$ through the inverse
Laplace transform,\begin{equation}
\rho(\mathcal{V};\phi)=\frac{1}{2\pi\text{i}}\int_{-\text{i}\infty}^{\text{i}\infty}\negmedspace dz\, e^{z\mathcal{V}}g(z;\phi),\label{eq:rho integral}\end{equation}
where the integration contour in the complex $z$ plane can be chosen
along the imaginary axis because all the singularities of $g(z;\phi)$
are to the left of that axis. The RV cutoff procedure depends on the
limit of $\rho(\mathcal{V};\phi)$ and related distributions at $\mathcal{V}\rightarrow\infty$.
The asymptotic behavior at $\mathcal{V}\rightarrow\infty$ is determined
by the type and the location of the right-most singularity of $g(z;\phi)$
in the half-plane $\mbox{Re}\, z<0$. For instance, if $z=z_{*}$
is such a singularity, the asymptotic is $\rho(\mathcal{V};\phi)\propto\exp[-z_{*}\mathcal{V}]$.
The prefactor in this expression needs to be determined; for this,
a detailed analysis of the singularities of $g(z;\phi)$ will be carried
out.

It is important to verify that the singularities of $g(z;\phi)$ are
$\phi$-independent. We first show that solutions of Eq.~(\ref{eq:g equ gen 1})
cannot diverge at finite values of $\phi$. If that were the case
and say $g(z;\phi)\rightarrow\infty$ as $\phi\rightarrow\phi_{1}$,
the function $\ln\ln g$ as well as derivatives $g_{,\phi}$ and $g_{,\phi\phi}$
would diverge as well. Then \begin{equation}
\lim_{\phi\rightarrow\phi_{1}}\partial_{\phi}\left[\ln\ln g\right]=\lim_{\phi\rightarrow\phi_{1}}\frac{\partial_{\phi}g}{g\ln g}=\infty.\end{equation}
It follows that the term $g\ln g$ is negligible near $\phi=\phi_{1}$
in Eq.~(\ref{eq:g equ gen 1}) compared with the term $\partial_{\phi}g$
and hence also with the term $\partial_{\phi}\partial_{\phi}g$. In
a very small neighborhood of $\phi=\phi_{1}$, the operator $\hat{L}$
can be approximated by a linear operator $\hat{L}_{1}$ with constant
coefficients, such as \begin{equation}
\hat{L}\approx\hat{L}_{1}\equiv A_{1}\partial_{\phi}\partial_{\phi}+B_{1}\partial_{\phi}.\end{equation}
Since at least one of the coefficients $A_{1},B_{1}$ is nonzero at
$\phi=\phi_{1}$, it follows that $g(z;\phi)$ is approximately a
solution of the linear equation $\hat{L}_{1}g=0$ near $\phi=\phi_{1}$.
However, solutions of linear equations cannot diverge at finite values
of the argument. Hence, the function $g(z;\phi)$ cannot diverge at
a finite value of $\phi$. 

The only remaining possibility is that the function $g(z;\phi)$ has
singular points $z=z_{*}$ such that $g(z_{*};\phi)$ remains finite
while $\partial g/\partial z$, or a higher-order derivative, diverges
at $z=z_{*}$. We will now investigate such divergences and show that
$g(z;\phi)$ has a leading singularity of the form\begin{equation}
g(z;\phi)=g(z_{*};\phi)+\sigma(\phi)\sqrt{z-z_{*}}+O(z-z_{*}),\label{eq:g sing structure}\end{equation}
where $z_{*}$ is a $\phi$-independent location of the singularity
such that $z_{*}\leq0$, while the function $\sigma(\phi)$ is yet
to be determined.

Denoting temporarily $g_{1}(z;\phi)\equiv\partial g/\partial z$,
we find a \emph{linear} equation for $g_{1}$,\begin{equation}
\hat{L}g_{1}+3\left(\ln g+1\right)g_{1}=0,\label{eq:g1 equ}\end{equation}
with inhomogeneous boundary conditions\begin{equation}
g_{1}(\phi_{*})=-H^{-3}(\phi_{*})e^{-zH^{-3}(\phi_{*})},\quad g_{1}(\phi_{\text{Pl}})=0.\label{eq:g1 bc}\end{equation}
The solution $g_{1}(z;\phi)$ of this linear problem can be found
using a standard method involving the Green's function. The problem
with inhomogeneous boundary conditions is equivalent to the problem
with zero boundary conditions but with an inhomogeneous equation.
To be definite, let us consider the operator $\hat{L}$ of the form
used in Eq.~(\ref{eq:g equ}),\begin{equation}
\hat{L}=\frac{D(\phi)}{H(\phi)}\partial_{\phi}\partial_{\phi}+\frac{v(\phi)}{H(\phi)}\partial_{\phi}.\label{eq:L one dim}\end{equation}
Then Eqs.~(\ref{eq:g1 equ})--(\ref{eq:g1 bc}) are equivalent to
the inhomogeneous problem with zero boundary conditions,\begin{align}
\hat{L}g_{1}+3\left(\ln g+1\right)g_{1} & =DH^{-4}e^{-zH^{-3}}\delta^{\prime}(\phi-\phi_{*}),\label{eq:g1 delta eq}\\
g_{1}(z;\phi_{*}) & =g_{1}(z;\phi_{\text{Pl}})=0.\label{eq:g1 delta bc}\end{align}
 The solution of this inhomogeneous equation exists as long as the
linear operator $\hat{L}+3(\ln g+1)$ does not have a zero eigenfunction
with zero boundary conditions.

Note that the operator $\hat{L}+3(\ln g+1)$ is explicitly $z$-dependent
through the coefficient $g(z;\phi)$. Note also that $g(z;\phi)\neq0$
by definition~(\ref{eq:g def}) for values of $z$ such that the
integral in Eq.~(\ref{eq:g def}) converges; hence $\ln g$ is finite
for those $z$. Let us denote by $G(z;\phi,\phi')$ the Green's function
of that operator with zero boundary conditions,\begin{align}
\hat{L}G+3\left(\ln g(z,\phi)+1\right)G & =\delta(\phi-\phi'),\\
G(z;\phi_{*},\phi') & =G(z;\phi_{\text{Pl}},\phi')=0.\end{align}
This Green's function is well-defined for values of $z$ such that
$\hat{L}+3(\ln g(z;\phi)+1)$ is invertible. For these $z$ we may
express the solution $g_{1}(z;\phi)$ of Eqs.~(\ref{eq:g1 equ})--(\ref{eq:g1 bc})
explicitly through the Green's function as\begin{equation}
g_{1}(z;\phi)=-\left.\frac{D}{H^{4}}e^{-zH^{-3}}\right|_{\phi_{*}}\left.\frac{\partial G(z;\phi,\phi')}{\partial\phi'}\right|_{\phi'=\phi_{*}}.\label{eq:g1 through Green}\end{equation}
Hence, for these $z$ the function $g_{1}(z;\phi)\equiv\partial g/\partial z$
remains finite at every value of $\phi$. A similar argument shows
that all higher-order derivatives $\partial^{n}g/\partial z^{n}$
remain finite at every $\phi$ for these $z$. Therefore, the singularities
of $g(z;\phi)$ can occur only at certain $\phi$-independent points
$z=z_{*}$, $z=z_{*}^{\prime}$, etc.

Since the generating function $g(z;\phi)$ is nonsingular for all
complex $z$ with $\mbox{Re}\, z>0$, it is assured that $g_{1}(z;\phi)$
and $G(z;\phi,\phi')$ exist for such $z$. However, there will be
values of $z$ for which the operator $\hat{L}+3(\ln g+1)$ has a
zero eigenfunction with zero boundary conditions, so the Green's function
$G$ is undefined. Denote by $z_{*}$ such a value with the algebraically
largest real part; we already know that $\mbox{Re}\, z_{*}\leq0$
in any case. Let us now show that the function $g_{1}(z;\phi)$ actually
diverges when $z\rightarrow z_{*}$. In other words, $\lim_{z\rightarrow z_{*}}g_{1}(z;\phi)=\infty$
for every value of $\phi$.

To show this, we need to use the decomposition of the Green's function
in the eigenfunctions of the operator $\hat{L}+3(\ln g+1)$,\begin{equation}
G(z;\phi,\phi')=\sum_{n=0}^{\infty}\frac{1}{\lambda_{n}(z)}f_{n}(\phi)f_{n}^{*}(\phi'),\label{eq:Green decomp}\end{equation}
where $f_{n}(z;\phi)$ are the (appropriately normalized) eigenfunctions
with eigenvalues $\lambda_{n}(z)$ and zero boundary conditions,\begin{align}
\left[\hat{L}+3(\ln g(z;\phi)+1)\right]f_{n}(z;\phi) & =\lambda_{n}(z)f_{n}(z;\phi),\label{eq:fn equ}\\
f_{n}(z;\phi)=0 & \,\:\mbox{for}\,\:\phi=\phi_{*},\,\phi=\phi_{\text{Pl}}.\label{eq:fn bc}\end{align}
The decomposition~(\ref{eq:Green decomp}) is possible as long as
the operator $\hat{L}$ is self-adjoint with an appropriate choice
of the scalar product in the space of functions $f(\phi)$. The scalar
product can be chosen in the following way,\begin{equation}
\left\langle f_{1},f_{2}\right\rangle =\int f_{1}(\phi)f_{2}^{*}(\phi)M(\phi)d\phi,\label{eq:f1 f2 scalar}\end{equation}
where $M(\phi)$ is a weighting function. One can attempt to determine
$M(\phi)$ such that the operator $\hat{L}$ is self-adjoint,\begin{equation}
\langle f_{1},\hat{L}f_{2}\rangle=\langle\hat{L}f_{1},f_{2}\rangle.\end{equation}
In single-field models of inflation where the operator $\hat{L}$
has the form~(\ref{eq:L one dim}), it is always possible to choose
$M(\phi)$ appropriately~\cite{Winitzki:1995pg}. However, in multi-field
models this is not necessarily possible.%
\footnote{I am grateful to D. Podolsky for pointing this out to me. The hermiticity
of operators of diffusion type in the context of eternal inflation
was briefly discussed in Ref.~\cite{Podolsky:2008du}.%
} One can show that in standard slow-roll models with $K$ fields $\phi_{1},...,\phi_{K}$
and kinetic coefficients\begin{equation}
D_{ij}=\frac{H^{3}}{8\pi^{2}}\delta_{ij},\quad v_{i}=-\frac{1}{4\pi G}\frac{\partial H}{\partial\phi_{i}},\quad H=H(\phi_{1},...,\phi_{K}),\end{equation}
there exists a suitable choice of $M(\phi)$, namely \begin{equation}
M(\phi_{1},...,\phi_{K})=\frac{\pi G}{H^{2}}\exp\left[\frac{\pi G}{H^{2}}\right],\end{equation}
 such that the operator \begin{equation}
\hat{L}=H^{-1}\sum_{i,j}D_{ij}\frac{\partial^{2}}{\partial\phi_{i}\partial\phi_{j}}+H^{-1}\sum_{i}v_{i}\frac{\partial}{\partial\phi_{i}}\end{equation}
is self-adjoint in the space of functions $f(\phi)$ with zero boundary
conditions and the scalar product~(\ref{eq:f1 f2 scalar}). However,
the operator $\hat{L}$ may be non-self-adjoint in more general inflationary
models where the kinetic coefficients are given by different expressions.
We omit the formulation of precise conditions for self-adjointness
of $\hat{L}$ because this property is not central to the present
investigation. In non-self-adjoint cases a decomposition similar to
Eq.~(\ref{eq:Green decomp}) needs to be performed using the left
and the right eigenfunctions of the non-self-adjoint operator $\hat{L}+3(\ln g+1)$.
One expects that such a decomposition will still be possible because
(heuristically) the nondiagonalizable operators are a set of measure
zero among all operators. The requisite left and right eigenfunctions
can be obtained numerically. We leave the detailed investigation of
those cases for future work. Presently, let us focus on the case when
the decomposition of the form~(\ref{eq:Green decomp}) holds, with
appropriately chosen scalar product and the normalized eigenfunctions\begin{equation}
\left\langle f_{m},f_{n}\right\rangle =\delta_{mn}.\end{equation}
The eigenfunctions $f_{m}(z;\phi)$ can be obtained e.g.~numerically
by solving the boundary value problem~(\ref{eq:fn equ})--(\ref{eq:fn bc}).

In the limit $z\rightarrow z_{*}$, one of the eigenvalues $\lambda_{n}$
approaches zero. Since linear operators such as $\hat{L}$ always
have a spectrum bounded from above~\cite{Winitzki:1995pg}, we may
renumber the eigenvalues $\lambda_{n}$ such that $\lambda_{0}$ is
the largest one. Then we define $z_{*}$ as the value with the (algebraically)
largest real part, such that $\lambda_{0}(z_{*})=0$. By construction,
for all $z$ with $\mbox{Re}\, z>\mbox{Re}\, z_{*}$ all the eigenvalues
$\lambda_{n}$ are negative. Note that the (algebraically) largest
eigenvalue $\lambda_{0}(z)$ is always nondegenerate, and the corresponding
eigenfunction $f_{0}(z;\phi)$ can be chosen real and positive for
all $\phi$, except at the boundaries $\phi=\phi_{*}$ and $\phi=\phi_{\text{Pl}}$
where $f$ satisfies the zero boundary conditions.

For $z$ near $z_{*}$, only the nondegenerate eigenvalue $\lambda_{0}$
will be near zero, so the decomposition~(\ref{eq:Green decomp})
of the Green's function will be dominated by the term $1/\lambda_{0}$.
Hence, we can use Eqs.~(\ref{eq:g1 through Green}) and (\ref{eq:Green decomp})
to determine the function $g_{1}(z;\phi)$ approximately as\begin{equation}
g_{1}(z;\phi)\approx-\frac{f_{0}(z;\phi)}{\lambda_{0}(z)}\frac{\partial f_{0}(z;\phi_{*})}{\partial\phi}\left[\frac{D}{H^{4}}e^{-zH^{-3}}\right]_{\phi_{*}}.\label{eq:g1 explicit}\end{equation}
It follows that indeed $g_{1}(z;\phi)\rightarrow\infty$ as $z\rightarrow z_{*}$
because $\lambda_{0}(z)\rightarrow0$. 

This detailed investigation allows us now to determine the behavior
of $g(z;\phi)$ at the leading singularity $z=z_{*}$. We will consider
the function $g(z;\phi)$ for $z$ near $z_{*}$ and show that the
singularity indeed has the structure~(\ref{eq:g sing structure}). 

We have already shown that the function $g$ itself does not diverge
at $z=z_{*}$ but its derivative $g_{1}\equiv\partial g/\partial z$
does. Hence, the function $g(z_{*};\phi)$ is continuous, and the
difference $g(z_{*};\phi)-g(z;\phi)$ is small for $z\approx z_{*}$,
so that we have the expansion \begin{align}
\delta g(z;\phi) & \equiv g(z;\phi)-g(z_{*};\phi)\\
 & \approx g_{1}(z;\phi)\left(z-z_{*}\right)+O\left[(z_{*}-z)^{2}\right].\label{eq:delta g g1}\end{align}
(Note that we are using the \emph{finite} value $g_{1}(z;\phi)$ rather
than the divergent value $g_{1}(z_{*};\phi)$ in the above equation.)
On the other hand, we have the explicit representation~(\ref{eq:g1 explicit}).
Let us examine the values of $\lambda_{0}(z)$ for $z\approx z_{*}$.
At $z=z_{*}$ we have $\lambda_{0}(z_{*})=0$, so the (small) value
$\lambda_{0}(z)$ for $z\approx z_{*}$ can be found using standard
perturbation theory for linear operators. If we denote the change
in the operator $\hat{L}$ by \begin{equation}
\delta\hat{L}\equiv3(\ln g(z;\phi)-\ln g(z_{*};\phi))\approx\frac{3\delta g(z;\phi)}{g(z_{*};\phi)},\end{equation}
we can write, to first order, \begin{equation}
\lambda_{0}(z)\approx\left\langle f_{0},\delta\hat{L}f_{0}\right\rangle =\int\left|f_{0}(\phi)\right|^{2}\frac{3\delta g(z;\phi)}{g(z_{*};\phi)}d\phi.\label{eq:lambda 0 integral}\end{equation}
Now, Eqs.~(\ref{eq:g1 explicit}) and (\ref{eq:delta g g1}) yield\begin{equation}
\frac{\delta g(z;\phi)}{z-z_{*}}\approx-\frac{f_{0}(z;\phi)}{\lambda_{0}(z)}\left[\frac{\partial f_{0}}{\partial\phi}\frac{D}{H^{4}}e^{-zH^{-3}}\right]_{\phi_{*}}.\end{equation}
Integrating the above equation in $\phi$ with the prefactor\begin{equation}
\left|f_{0}(\phi)\right|^{2}\frac{3}{g(z_{*};\phi)}d\phi\end{equation}
and using Eq.~(\ref{eq:lambda 0 integral}), we obtain a closed equation
for $\lambda_{0}(z)$ in which terms of order $\left(z-z_{*}\right)^{2}$
have been omitted, \begin{align}
\frac{\lambda_{0}(z)}{z-z_{*}} & \approx-\frac{1}{\lambda_{0}(z)}\left[\frac{\partial f_{0}}{\partial\phi}\frac{D}{H^{4}}e^{-zH^{-3}}\right]_{\phi_{*}}\nonumber \\
 & \quad\times\int\left|f_{0}\right|^{2}\frac{3\delta g(z;\phi)}{g(z_{*};\phi)}f_{0}(\phi)d\phi.\label{eq:lambda 0 ans}\end{align}
It follows that $\lambda_{0}(z)\propto\sqrt{z-z_{*}}$ and $g_{1}(z;\phi)\propto\left(z-z_{*}\right)^{-1/2}$,
confirming the leading asymptotic of the form~(\ref{eq:g sing structure}).

Let us also obtain a more explicit form of the singularity structure
of $g(z;\phi)$. We can rewrite Eq.~(\ref{eq:lambda 0 ans}) as\begin{equation}
\lambda_{0}(z)\approx\sigma_{0}\sqrt{z-z_{*}}+O(z-z_{*}),\label{eq:lambda 0 sing}\end{equation}
 where $\sigma_{0}$ is a constant that may be obtained explicitly.
Then Eq.~(\ref{eq:g1 explicit}) yields\begin{equation}
g_{1}(z;\phi)\approx\frac{f_{0}(z;\phi)}{\sqrt{z-z_{*}}}\sigma_{1},\end{equation}
with a different constant $\sigma_{1}$. Finally, we can integrate
this in $z$ and obtain\begin{equation}
g(z;\phi)=g(z_{*};\phi)+2\sigma_{1}f_{0}(z;\phi)\sqrt{z-z_{*}}+O(z-z_{*}).\end{equation}
We may rewrite this by substituting $z=z_{*}$ into $f_{0}(z;\phi)$,\begin{align}
g(z;\phi) & =g(z_{*};\phi)+\sigma(\phi)\sqrt{z-z_{*}}+O(z-z_{*}),\label{eq:g asympt sigma 1}\\
\sigma(\phi) & \equiv2\sigma_{1}f_{0}(z_{*};\phi).\end{align}
The result is now explicitly of the form~(\ref{eq:g sing structure}).
It will turn out that the normalization constant $2\sigma_{1}$ cancels
in the final results. So in a practical calculation the eigenfunction
$f_{0}(z_{*};\phi)$ may be determined with an arbitrary normalization. 

As a side note, let us remark that the argument given above will apply
also to other singular points $z_{*}^{\prime}\neq z_{*}$ as long
as the eigenvalue $\lambda_{k}(z)$ of the operator $\hat{L}+3(\ln g+1)$
is nondegenerate when it vanishes at $z=z_{*}^{\prime}$. If the relevant
eigenvalue becomes degenerate, the singularity structure will not
be of the form $\sqrt{z-z_{*}^{\prime}}$ but rather $\left(z-z_{*}^{\prime}\right)^{s}$
with some other power $0<s<1$.

Now we are ready to obtain the asymptotic form of the distribution
$\rho(\mathcal{V};\phi)$ for $\mathcal{V}\rightarrow\infty$. We
deform the integration contour in the inverse Laplace transform~(\ref{eq:rho integral})
such that it passes near the real axis around $z=z_{*}$. Then we
use Eq.~(\ref{eq:g asympt sigma 1}) for $g(z;\phi)$ and obtain
the leading asymptotic \begin{align}
\rho(\mathcal{V};\phi) & \approx\frac{1}{2\pi\text{i}}\sigma(\phi)\left[\int_{-\infty}^{z_{*}}-\int_{z_{*}}^{-\infty}\right]\sqrt{z-z_{*}}\, e^{z\mathcal{V}}dz\nonumber \\
 & =\frac{1}{2\sqrt{\pi}}\sigma(\phi)\mathcal{V}^{-3/2}e^{z_{*}\mathcal{V}}.\label{eq:rho asympt 1}\end{align}
The subdominant terms come from the higher-order terms in the expansion
in Eq.~(\ref{eq:g asympt sigma 1}) and are of the order $\mathcal{V}^{-1}$
times the leading term shown in Eq.~(\ref{eq:rho asympt 1}).

Finally, we show that $z_{*}$ must be real-valued and that there
are no other singularities $z_{*}^{\prime}$ with $\mbox{Re}\, z_{*}^{\prime}=\mbox{Re}\, z_{*}$.
This is so because the integral\begin{equation}
g_{1}(z_{*};\phi)=-\int_{0}^{\infty}\rho(\mathcal{V};\phi)e^{-z_{*}\mathcal{V}}\mathcal{V}d\mathcal{V}=\infty\label{eq:g1 integral}\end{equation}
will definitely diverge for purely real $z_{*}$ if it diverges for
a nonreal value $z_{*}^{\prime}=z_{*}+\text{i}A$. If, on the other
hand, the integral~(\ref{eq:g1 integral}) diverges for a real $z_{*}$,
it will \emph{converge} for any nonreal $z_{*}^{\prime}=z_{*}+\text{i}A$
with real $A\neq0$ because the function $\rho(\mathcal{V};\phi)$
has the large-$\mathcal{V}$ asymptotic of the form~(\ref{eq:rho asympt 1})
and the oscillations of $\exp(\text{i}A\mathcal{V})$ will make the
integral~(\ref{eq:g1 integral}) convergent.

\subsection{FPRV distribution of a field $Q$\label{sub:FPRV-distribution-of-Q}}

In this section we follow the notation of Ref.~\cite{Winitzki:2008yb}.

Consider a fluctuating field $Q$ such that the Fokker-Planck operator
$\hat{L}$ is of the form~(\ref{eq:FP operator L for Q}). We are
interested in the portion $\mathcal{V}_{Q_{R}}$ of the total reheated
volume $\mathcal{V}$ where the field $Q$ has a value within a given
interval $\left[Q_{R},Q_{R}+dQ\right]$. We denote by $\rho(\mathcal{V},\mathcal{V}_{Q_{R}};\phi_{0},Q_{0})$
the joint probability distribution of the volumes $\mathbf{\mathcal{V}}$
and $\mathcal{V}_{Q_{R}}$ for initial $H$-regions with initial values
$\phi=\phi_{0}$ and $Q=Q_{0}$. The generating function $\tilde{g}(z,q;\phi,Q)$
corresponding to that distribution is defined by\begin{equation}
\tilde{g}(z,q;\phi,Q)\equiv\int e^{-z\mathcal{V}-q\mathcal{V}_{Q}}\rho(\mathcal{V},\mathcal{V}_{Q};\phi,Q)d\mathcal{V}d\mathcal{V}_{Q}.\end{equation}
Since this generating function satisfies the same multiplicative property~(\ref{eq:g multiplicative})
as the generating function $g(z;\phi,Q)$, we may repeat the derivation
of Eq.~(\ref{eq:g equ gen 1}) without modifications for the function
$\tilde{g}(z,q;\phi,Q)$. Hence, $g(z,q;\phi,Q)$ is the solution
of the same equation as $g(z;\phi,Q)$. The only difference is the
boundary conditions at reheating, which are given not by Eq.~(\ref{eq:bc reheating phi Q})
but by\begin{equation}
\tilde{g}(z,q;\phi_{*},Q)=\exp\left[-\left(z+q\delta_{QQ_{R}}\right)H^{-3}(\phi_{*},Q)\right],\label{eq:bc for g tilda z q}\end{equation}
where (with a slight abuse of notation) $\delta_{QQ_{R}}$ is the
indicator function of the interval $\left[Q_{R},Q_{R}+dQ\right]$,
i.e.\begin{equation}
\delta_{QQ_{R}}\equiv\theta(Q-Q_{R})\theta(Q_{R}+dQ-Q).\end{equation}
We employ this {}``finite'' version of the $\delta$-function only
because we cannot use a standard Dirac $\delta$-function under the
exponential. This slight technical inconvenience will disappear shortly.

The solution for the function $\tilde{g}(z,q;\phi,Q)$ may be obtained
in principle and will provide complete information about the distribution
of possible values of the volume $\mathcal{V}_{Q}$ together with
the total reheating volume $\mathcal{V}$ to the future of an initial
$H$-region. In the context of the RV cutoff, one is interested in
the event when $\mathcal{V}$ is finite and very large. Then one expects
that $\mathcal{V}_{Q}$ also becomes typically very large while the
ratio $\mathcal{V}_{Q}/\mathcal{V}$ remains roughly constant. In
other words, one expects that the distribution of $\mathcal{V}_{Q}$
is sharply peaked around a mean value $\left\langle \mathcal{V}_{Q}\right\rangle $,
and that the limit $\left\langle \mathcal{V}_{Q}\right\rangle /\mathcal{V}$
is well-defined at $\mathcal{V}\rightarrow\infty$. The value of that
limit is the only information we need for calculations in the RV cutoff.
Therefore, we do not need to compute the entire distribution $\rho(\mathcal{V},\mathcal{V_{Q}};\phi,Q)$
but only the mean value $\left\langle \left.\mathcal{V}_{Q}\right|_{\mathcal{V}}\right\rangle $
at fixed $\mathcal{V}$. 

Let us therefore define the generating function of the mean value
$\left\langle \left.\mathcal{V}_{Q}\right|_{\mathcal{V}}\right\rangle $
as follows,\begin{equation}
h(z;\phi,Q)\equiv\left\langle \mathcal{V}_{Q_{R}}e^{-z\mathcal{V}}\right\rangle _{\mathcal{V}<\infty}=-\frac{\partial\tilde{g}}{\partial q}(z,q=0;\phi,Q).\end{equation}
(The dependence on the fixed value of $Q_{R}$ is kept implicit in
the function $h(z;\phi,Q)$ in order to make the notation less cumbersome.)
The differential equation and the boundary conditions for $h(z;\phi,Q)$
follow straightforwardly by taking the derivative $\partial_{q}$
at $q=0$ of Eqs.~(\ref{eq:g equ gen 1}) and (\ref{eq:bc for g tilda z q}).
It is clear from the definition of $\tilde{g}$ that $\tilde{g}(z,q=0;\phi,Q)=g(z;\phi,Q)$.
Hence we obtain\begin{align}
\hat{L}h+3\left(\ln g(z;\phi,Q)+1\right)h & =0,\label{eq:h equ pre 0}\\
h(z;\phi_{*},Q) & =\frac{e^{-zH^{-3}(\phi_{*},Q)}}{H^{3}(\phi_{*},Q)}\delta_{QQ_{R}},\\
h(z;\phi_{\text{Pl}},Q) & =0.\end{align}
Note that it is the generating function $g$, not $\tilde{g}$, that
appears as a coefficient in Eq.~(\ref{eq:h equ pre 0}). 

Since the {}``finite'' $\delta$-function $\delta_{QQ_{R}}$ now
enters only linearly rather than under an exponential, we may replace
$\delta_{QQ_{R}}$ by the ordinary Dirac $\delta$-function $\delta(Q-Q_{R})$.
To maintain consistency, we need to divide $h$ by $dQ$, which corresponds
to computing the \emph{probability} \emph{density} of the reheated
volume with $Q=Q_{R}$. This probability density is precisely the
goal of the present calculation. 

The RV-regularized probability density for values of $Q$ is defined
as the limit\begin{align}
p(Q_{R}) & =\lim_{\mathcal{V}\rightarrow\infty}\frac{\left\langle \left.\mathcal{V}_{Q_{R}}\right|_{\mathcal{V}}\right\rangle _{\mathcal{V}<\infty}}{\mathcal{V}\,\rho(\mathcal{V};\phi,Q)}\nonumber \\
 & =\lim_{\mathcal{V}\rightarrow\infty}\frac{\int_{-\text{i}\infty}^{\text{i}\infty}\! e^{z\mathcal{V}}h(z;\phi,Q)dz}{\mathcal{V}\int\! e^{z\mathcal{V}}g(z;\phi,Q)dz}.\label{eq:pQR limit 1}\end{align}
To compute this limit, we need to consider the asymptotic behavior
of $\left\langle \left.\mathcal{V}_{Q_{R}}\right|_{\mathcal{V}}\right\rangle _{\mathcal{V}<\infty}$
at $\mathcal{V}\rightarrow\infty$. This behavior is determined by
the leading singularity of the function $h(z;\phi,Q)$ in the complex
$z$ plane. The arguments of Sec.~\ref{sub:Singularities-of-g} apply
also to $h(z;\phi,Q)$ and show that $h$ cannot have a $\phi$- or
$Q$-dependent singularity in the $z$ plane. 

Moreover, $h(z;\phi,Q)$ has precisely the same singular points, in
particular $z=z_{*}$, as the basic generating function $g(z;\phi,Q)$
of the reheating volume. Indeed, the function $h(z;\phi,Q)$ can be
expressed through the Green's function $G(z;\phi,Q,\phi',Q')$ of
the operator $\hat{L}+3(\ln g+1)$, similarly to the function $g_{1}(z;\phi)$
considered in Sec.~\ref{sub:Singularities-of-g}. For $z\neq z_{*}$,
this operator is invertible on the space of functions $f(\phi,Q)$
satisfying zero boundary conditions. Hence, $h(z;\phi,Q)$ is nonsingular
at $z\neq z_{*}$ and becomes singular precisely at $z=z_{*}$.

Let us now obtain an explicit form of $h(z;\phi,Q)$ near the singular
point $z=z_{*}$. We assume again the eigenfunction decomposition
of the Green's function (with the same caveats as in Sec.~\ref{sub:Singularities-of-g}),\begin{equation}
G(z;\phi,Q,\phi',Q')=\sum_{n=0}^{\infty}\frac{1}{\lambda_{n}(z)}f_{n}(z;\phi,Q)f_{n}^{*}(z;\phi',Q'),\end{equation}
where $f_{n}$ are appropriately normalized eigenfunctions of the
$z$-dependent operator $\hat{L}+3(\ln g+1)$ with eigenvalues $\lambda_{n}(z)$.
The eigenfunctions $f_{n}$ must satisfy zero boundary conditions
at reheating and Planck boundaries. Similarly to the way we derived
Eq.~(\ref{eq:g1 explicit}), we obtain the explicit solution\begin{equation}
h(z;\phi,Q)=\sum_{n=0}^{\infty}\frac{f_{n}(z;\phi,Q)}{\lambda_{n}(z)}\left[\frac{\partial f_{n}}{\partial\phi}\frac{D_{\phi\phi}e^{-zH^{-3}}}{H^{4}}\right]_{\phi_{*},Q_{R}}.\end{equation}
The value of $h(z;\phi,Q)$ for $z\approx z_{*}$ is dominated by
the contribution of the large factor $1/\lambda_{0}(z)\propto\left(z-z_{*}\right)^{-1/2}$,
so the leading term is \begin{equation}
h(z;\phi,Q)\approx\frac{f_{0}(z_{*};\phi,Q)}{\lambda_{0}(z)}\left[\frac{\partial f_{0}}{\partial\phi}\frac{D_{\phi\phi}e^{-z_{*}H^{-3}}}{H^{4}}\right]_{\phi_{*},Q_{R}}.\end{equation}
The asymptotic behavior of the mean value $\left\langle \left.\mathcal{V}_{Q_{R}}\right|_{\mathcal{V}}\right\rangle _{\mathcal{V}<\infty}$
is determined by the singularity of $h(z;\phi,Q)$ at $z=z_{*}$.
As before, we may deform the integration contour to pass near the
real axis around $z=z_{*}$. We can then express the large-$\mathcal{V}$
asymptotic of $\left\langle \left.\mathcal{V}_{Q_{R}}\right|_{\mathcal{V}}\right\rangle _{\mathcal{V}<\infty}$
as follows,\begin{align}
\left\langle \left.\mathcal{V}_{Q_{R}}\right|_{\mathcal{V}}\right\rangle  & =\frac{1}{2\pi\text{i}}\int_{-\text{i}\infty}^{\text{i}\infty}\! e^{z\mathcal{V}}h(z;\phi,Q)dz\nonumber \\
 & \approx\frac{f_{0}(z_{*};\phi,Q)}{\sqrt{\pi}\sigma_{0}\sqrt{\mathcal{V}}}e^{z_{*}\mathcal{V}}\left[\frac{\partial f_{0}}{\partial\phi}\frac{D_{\phi\phi}e^{-z_{*}H^{-3}}}{H^{4}}\right]_{\phi_{*},Q_{R}},\label{eq:V QR ave ans}\end{align}
where $\sigma_{0}$ is the constant defined by Eq.~(\ref{eq:lambda 0 sing}).

We now complete the analytic evaluation of the limit~(\ref{eq:pQR limit 1}).
Since the denominator of Eq.~(\ref{eq:pQR limit 1}) has the large-$\mathcal{V}$
asymptotics of the form\begin{equation}
\mathcal{V}\int_{0}^{\infty}\negmedspace g(z;\phi,Q)e^{z\mathcal{V}}d\mathcal{V}\propto f_{0}(z_{*};\phi,Q)\mathcal{V}^{-\frac{1}{2}}e^{z_{*}\mathcal{V}},\end{equation}
where $f_{0}$ is the same eigenfunction, the dependence on $\phi$
and $Q$ identically cancels in the limit~(\ref{eq:pQR limit 1}).
Hence, that limit is independent of the initial values $\phi$ and
$Q$ but is a function only of $Q_{R}$, on which $h(z;\phi,Q)$ implicitly
depends. Using this fact,, we can significantly simplify the rest
of the calculation. It is not necessary to compute the denominator
of Eq.~(\ref{eq:pQR limit 1}) explicitly. The distribution of the
values of $Q$ at $\phi=\phi_{*}$ is simply proportional to the $Q_{R}$-dependent
part of Eq.~(\ref{eq:V QR ave ans}); the denominator of Eq.~(\ref{eq:pQR limit 1})
serves merely to normalize that distribution. Hence, the RV cutoff
yields\begin{equation}
p(Q_{R})=\mbox{const}\left[\frac{\partial f_{0}(z_{*};\phi,Q)}{\partial\phi}\frac{D_{\phi\phi}e^{-z_{*}H^{-3}}}{H^{4}}\right]_{\phi=\phi_{*},Q=Q_{R}},\label{eq:PQR ans 1}\end{equation}
where the normalization constant needs to be chosen such that $\int p(Q_{R})dQ_{R}=1$.
This is the final analytic formula for the RV cutoff applied to the
distribution of $Q$ at reheating. The the value $z_{*}$, and the
corresponding solution $g(z_{*};\phi,Q)$ of Eq.~(\ref{eq:g equ 0}),
and the eigenfunction $f_{0}(z_{*};\phi,Q)$ need to be obtained numerically
unless an analytic solution is possible.

Let us comment on the presence of the factor $D_{\phi\phi}$ in the
formula~(\ref{eq:PQR ans 1}). The {}``diffusion'' coefficient
$D_{\phi\phi}$ is evaluated at the reheating boundary and is thus
small since the fluctuation amplitude at reheating is (in slow-roll
inflationary models)\begin{equation}
\frac{\delta\phi}{\phi}\sim\frac{H^{2}}{\dot{\phi}}=\frac{\sqrt{8\pi^{2}D_{\phi\phi}H}}{v_{\phi}}\sim10^{-5}.\end{equation}
Nevertheless it is not possible to set $D_{\phi\phi}=0$ directly
in Eq.~(\ref{eq:PQR ans 1}). This is so because the existence of
the Green's function of the Fokker-Planck operator such as $\hat{L}$
depends on the fact that $\hat{L}$ is a second-order differential
operator of elliptic type. If one sets $D_{\phi\phi}=0$ near the
reheating boundary, the operator $\hat{L}$ becomes first-order in
$\phi$ at that boundary. Then one needs to use a different formula
than Eq.~(\ref{eq:g1 delta eq}) for reducing an equation with inhomogeneous
boundary conditions to an inhomogeneous equation with zero boundary
conditions. Accordingly, one cannot use formulas such as Eq.~(\ref{eq:g1 through Green})
for the solutions. Alternative ways of solving the relevant equations
in that case will be used in Sec.~\ref{sub:A-model-of-slow-roll}.

\subsection{Calculations for an inflationary model\label{sub:A-model-of-slow-roll}}

In this section we perform explicit calculations of RV cutoff for
a model of slow-roll inflation driven by a scalar field with a potential
shown in Fig.~\ref{cap:pot1}. The kinetic coefficients $D(\phi)$
and $v(\phi)$ are such that $D(\phi)=D_{0}$, $v(\phi)=0$, and $H(\phi)=H_{0}$
in the flat region $\phi_{1}<\phi<\phi_{2}$, where the constants
$D_{0}$ and $H_{0}$ are \begin{equation}
H_{0}=\sqrt{\frac{8\pi G}{3}V_{0}},\quad D_{0}=\frac{H_{0}^{3}}{8\pi^{2}}.\end{equation}
In the slow-roll regions $\phi_{1}<\phi<\phi_{*}^{(1)}$ and $\phi_{2}<\phi<\phi_{*}^{(2)}$,
the coefficient $D(\phi)$ is set equal to zero, while $v(\phi)\ne0$
and $H(\phi)$ is not constant any more. The number of $e$-folds
in the two slow-roll {}``shoulders'' can be computed by the standard
formula, \begin{equation}
N_{j}=\int_{\phi_{j}}^{\phi_{*}^{(j)}}\frac{H}{v}d\phi=-4\pi G\int_{\phi_{j}}^{\phi_{*}^{(j)}}\frac{H}{H'}d\phi,\quad j=1,2.\end{equation}

\begin{figure}
\begin{centering}\psfrag{a}{$-\frac{6\pi^2}{H_0^2}$} \psfrag{U}{$U(g)$} \psfrag{0}{$0$} \psfrag{0.5}{$0.5$} \psfrag{1}{$1$} \psfrag{1.5}{$1.5$} \psfrag{g}{$g$} \psfrag{u0}{$E$} \psfrag{g0}{$g_0$} \psfrag{g1}{$g(z;\phi_1)$}  \psfrag{g2}{$g(z;\phi_2)$} \includegraphics[width=0.5\textwidth]{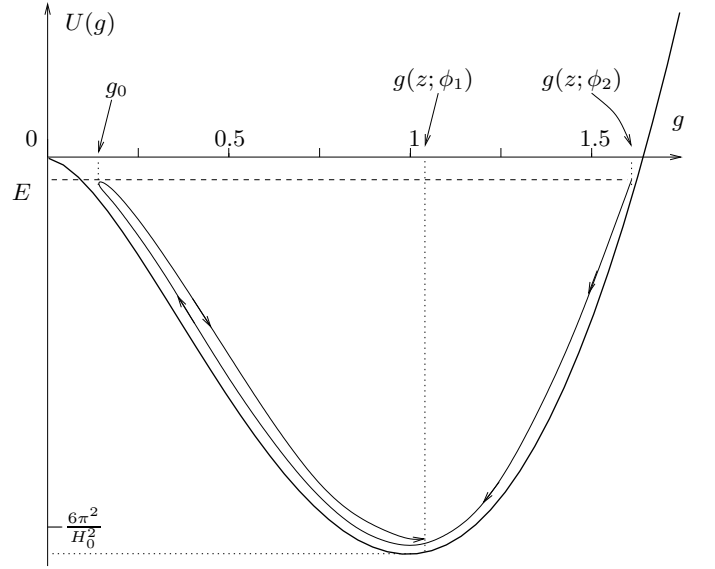}\par\end{centering}

\caption{Potential $U(g)$ given by Eq.~(\ref{eq:U pot def}) can be used
to interpret solutions $g(z;\phi)$ as mechanical motion in {}``time''
$\phi$ at constant total energy $E$. The potential vanishes at $g=0$
and $g=e^{1/2}$ and has a minimum at $g=1$. The value $g_{0}$ is
the turning point where $U(g_{0})=E$. A trajectory corresponding
to $z<0$ will begin and end with $g>1$, i.e.~to the right of the
minimum of the potential. Solutions cease to exist when $z<z_{*}$;
the solution $g(z_{*};\phi)$, shown by the thin line with arrows,
corresponds to a trajectory that starts at rest (as demonstrated in
the text). The energy of this trajectory is $E\approx0$, and so the
value $g(z_{*};\phi_{2})$ is close to $e^{1/2}$ while $g(z_{*};\phi_{1})$
is close to 1. \label{fig:Potential-U}}
\end{figure}

The first step of the calculation is to determine the singular point
$z=z_{*}$ of solutions $g(z;\phi)$ of Eq.~(\ref{eq:g equ}). We
expect $z_{*}$ to be real and negative. The boundary conditions for
$g(z;\phi)$ are\begin{equation}
g(z;\phi_{*}^{(1,2)})=\exp\left[-zH^{-3}(\phi_{*}^{(1,2)})\right].\label{eq:bc for g 12}\end{equation}
In each of the two deterministic regions, $\phi_{*}^{(1)}<\phi<\phi_{1}$
and $\phi_{2}<\phi<\phi_{*}^{(2)}$, Eq.~(\ref{eq:g equ}) becomes\begin{equation}
\frac{v}{H}\partial_{\phi}g+3g\ln g=0,\end{equation}
with the general solution\begin{equation}
g(z;\phi)=\exp\left[C\exp\left(-3\int^{\phi}\!\frac{H}{v}d\phi\right)\right],\end{equation}
where $C$ is an integration constant. Since the equation is first-order
within the deterministic regions, the solutions are fixed by the boundary
condition~(\ref{eq:bc for g 12}) in the respective region,\begin{equation}
g(z;\phi)=\exp\left[-zH^{-3}(\phi_{*}^{(1,2)})\exp\!\left(\!-3\!\int_{\phi_{*}^{(1,2)}}^{\phi}\!\!\frac{H}{v}d\phi\right)\!\right].\label{eq:g sol nodiff 12}\end{equation}
We may therefore compute the values of $g(z;\phi)$ at the boundaries
$\phi_{1,2}$ of the self-reproduction region as\begin{equation}
g(z;\phi_{1,2})=\exp\left[-zH^{-3}(\phi_{*}^{(1,2)})\exp\left(3N_{1,2}\right)\right].\label{eq:bc for g equ const 0}\end{equation}
Now we need to solve Eq.~(\ref{eq:g equ}) with these boundary conditions
in the region $\phi_{1}<\phi<\phi_{2}$. The equation has then the
form\begin{equation}
\frac{D_{0}}{H_{0}}\partial_{\phi}\partial_{\phi}g+3g\ln g=0.\label{eq:g equ const 0}\end{equation}
Exact solutions of Eq.~(\ref{eq:g equ const 0}) were studied in
Ref.~\cite{Winitzki:2001np}, to which the reader is referred for
more details. It is easy to show that Eq.~(\ref{eq:g equ const 0})
is formally equivalent to a one-dimensional motion of a particle with
coordinate $g(\phi)$ in a potential $U(g)$,\begin{equation}
U(g)=\frac{6\pi^{2}}{H_{0}^{2}}g^{2}\left(2\ln g-1\right),\label{eq:U pot def}\end{equation}
while $\phi$ plays the role of time. A solution $g(z;\phi)$ with
boundary conditions~(\ref{eq:bc for g equ const 0}) corresponds
to a trajectory that starts at the given value $g(z;\phi_{1})$ with
the initial velocity chosen such that the motion takes precisely the
specified time interval $\phi_{2}-\phi_{1}$ and reaches $g(z;\phi_{2})$.
For $z<0$ the boundary conditions specify $g(z;\phi_{1,2})>1$, i.e.~the
trajectory begins and ends to the right of the minimum of the potential
(see Fig.~\ref{fig:Potential-U}). Since the system is conservative,
there is a constant of motion $E$ (the {}``energy'') such that
$E=U(g_{0})$ at the highest point of the trajectory $g_{0}$ where
the {}``kinetic energy'' vanishes. The solution $g(z;\phi)$ can
be written implicitly as one of the two alternative formulas,\begin{equation}
\pm\int_{g(z;\phi)}^{g(z;\phi_{1,2})}\frac{dg}{\sqrt{2E(z)-2U(g)}}=\phi-\phi_{1,2},\label{eq:g sol implicit}\end{equation}
valid in appropriate intervals $\phi_{1}<\phi<\phi_{0}$ and $\phi_{0}<\phi<\phi_{2}$
respectively, where $\phi_{0}$ is the value of $\phi$ corresponding
to the turning point $g_{0}=g(z;\phi_{0})$. The value $E=E(z)$ in
Eq.~(\ref{eq:g sol implicit}) must be chosen such that the total
{}``time'' is $\phi_{2}-\phi_{1}$,\begin{equation}
\left[\int_{g_{0}}^{g(z;\phi_{1})}+\int_{g_{0}}^{g(z;\phi_{2})}\right]\frac{dg}{\sqrt{2E-2U(g)}}=\phi_{2}-\phi_{1}.\end{equation}
This condition together with $E=U(g_{0})$ implicitly determine the
values $E=E(z)$ and $g_{0}=g_{0}(z)$. 

The singularity $z=z_{*}$ of the solution $g(z;\phi)$ is found by
using the condition $\partial g/\partial z\rightarrow\infty$. Differentiating
Eq.~(\ref{eq:g sol implicit}) with respect to $z$ and substituting
Eq.~(\ref{eq:bc for g equ const 0}) for $g(z;\phi_{1,2})$, we obtain
the condition\begin{align}
 & -\frac{\partial g}{\partial z}\frac{1}{\sqrt{2E(z)-2U(g)}}+\frac{e^{3N_{1,2}}H^{-3}(\phi_{*}^{(1,2)})}{\sqrt{2E(z)-2U(g(z;\phi_{1,2}))}}\nonumber \\
 & -E^{\prime}(z)\int_{g(z;\phi)}^{g(z;\phi_{1,2})}\frac{dg}{\left[2E(z)-2U(g)\right]^{3/2}}=0.\end{align}
It follows that $\partial g/\partial z\rightarrow\infty$ when \begin{equation}
E(z)=U(g(z;\phi_{1,2})).\end{equation}
This condition is interpreted in the language of Fig.~\ref{fig:Potential-U}
as follows. As the value of $z$ becomes more negative, the initial
and the final values of $g$ given by Eq.~(\ref{eq:bc for g equ const 0})
both grow. The last available trajectory starts from rest at $\phi=\phi_{2}$
and at the value of $g$ such that $U(g)=E$. 

To obtain a specific result, let us assume that $\phi_{2}-\phi_{1}$
is sufficiently large to provide self-reproduction ($\phi_{2}-\phi_{1}\gg H_{0}$)
and that the number of $e$-folds in channel 1 is smaller than that
in channel 2, \begin{equation}
H^{-3}(\phi_{*}^{(1)})\exp\left(3N_{1}\right)\ll H^{-3}(\phi_{*}^{(2)})\exp\left(3N_{2}\right).\end{equation}
Then the value $g(z;\phi_{2})$ will grow faster than $g(z;\phi_{1})$
as $z$ becomes more negative. It follows that $g(z;\phi_{2})$ will
reach the singular point first. Since the {}``time'' $\phi_{2}-\phi_{1}$
is large, the constant $E$ will be close to 0 so that the trajectory
spends a long {}``time'' near $g=0$. Then the value $g(z_{*};\phi_{2})$
will be close to $e^{1/2}$. Hence the value of $z_{*}$ is approximately\begin{equation}
z_{*}\approx-\frac{1}{2}H^{3}(\phi_{*}^{(2)})\exp\left(-3N_{2}\right).\label{eq:z star estimate}\end{equation}
For this value of $z_{*}$, the starting point of the trajectory will
be \begin{equation}
g(z_{*};\phi_{1})\approx\exp\left[\frac{1}{2}\exp\left(3N_{1}-3N_{2}\right)\right]\approx1.\end{equation}
Hence, the solution $g(z_{*};\phi)$ at the singular point $z=z_{*}$
can be visualized as the thin line in Fig.~\ref{fig:Potential-U},
starting approximately at $g(z_{*};\phi_{1})=1$ and finishing at
$g(z_{*};\phi_{2})\approx e^{1/2}$.

An approximate expression for $g(z_{*};\phi)$ can be obtained by
setting $E\approx0$ in Eq.~(\ref{eq:g sol implicit}); then the
integral can be evaluated analytically. In the range $\phi_{0}<\phi<\phi_{2}$
we obtain \begin{equation}
\phi_{2}-\phi\approx\negmedspace\int_{g(z_{*};\phi)}^{g(z_{*};\phi_{2})}\negmedspace\frac{dg}{\sqrt{-2U(g)}}\approx\frac{H_{0}}{\sqrt{12\pi^{2}}}\left.\sqrt{1-2\ln g}\right|_{\phi_{2}}^{\phi},\end{equation}
so the solution is\begin{equation}
g(z_{*};\phi)\approx\exp\left[\frac{1}{2}-\frac{6\pi^{2}}{H_{0}^{2}}\left(\phi_{2}-\phi\right)^{2}\right],\quad\phi_{0}<\phi<\phi_{2}.\label{eq:g approx 2}\end{equation}
In the range $\phi_{1}<\phi<\phi_{0}$ we obtain within the same approximation\begin{equation}
g(z_{*};\phi)\approx\exp\left[\frac{1}{2}-\frac{6\pi^{2}}{H_{0}^{2}}\left(\phi-\phi_{1}+{\textstyle \frac{H_{0}}{\sqrt{12\pi^{2}}}}\right)^{2}\right].\label{eq:g approx 1}\end{equation}
These approximations are valid for $\phi$ within the indicated ranges
and away from the turning point $\phi_{0}$. The value of $\phi_{0}$
can be estimated by requiring that the value of $g(z_{*};\phi_{0})$
obtained from Eq.~(\ref{eq:g approx 2}) be equal to that obtained
from Eq.~(\ref{eq:g approx 1}). This yields\begin{equation}
\phi_{0}\approx\frac{1}{2}\left[\phi_{2}+\phi_{1}-{\textstyle \frac{H_{0}}{\sqrt{12\pi^{2}}}}\right]\approx\frac{\phi_{2}+\phi_{1}}{2}.\end{equation}
We note that the value $g(z_{*};\phi_{0})$ can be obtained somewhat
more precisely by approximating the solution $g(z_{*};\phi)$ in a
narrow interval near $\phi=\phi_{0}$ by a function of the form $\exp\left[A+B(\phi-\phi_{0})^{2}\right]$
and matching both the values and the derivatives of $g(z_{*};\phi)$
to the approximations~(\ref{eq:g approx 2}) and (\ref{eq:g approx 1})
at some intermediate points straddling $\phi=\phi_{0}$. In this way,
a uniform analytic approximation for $g(z_{*};\phi)$ can be obtained.
However, the accuracy of the approximations~(\ref{eq:g approx 2})
and (\ref{eq:g approx 1}) is sufficient for the present purposes.

Having obtained adequate analytic approximations for $z_{*}$ and
$g(z_{*},\phi)$, we can now proceed to the calculation of the mean
volumes $\left\langle \left.\mathcal{V}_{1,2}\right|_{\mathcal{V}}\right\rangle $
of regions reheated through channels 1 and 2 respectively, conditioned
on the event that the total volume of all reheated regions is $\mathcal{V}$.
We use the formalism developed in Sec.~\ref{sub:FPRV-distribution-of-Q},
where the variable $Q$ now takes only the discrete values 1 and 2,
so instead let us denote that value by $j$. The relevant generating
function $h_{j}(z;\phi)$ is defined by \begin{equation}
h_{j}(z;\phi)\equiv\left\langle \mathcal{V}_{j}e^{-z\mathcal{V}}\right\rangle _{\mathcal{V}<\infty},\quad j=1,2,\end{equation}
and is found as the solution of Eq.~(\ref{eq:h equ pre 0}), which
now takes the form\begin{equation}
\left[\frac{D(\phi)}{H(\phi)}\partial_{\phi}\partial_{\phi}+\frac{v(\phi)}{H(\phi)}\partial_{\phi}+3(\ln g(z;\phi)+1)\right]h_{j}(z;\phi)=0,\label{eq:h equ toy}\end{equation}
with boundary conditions imposed at the reheating boundaries,\begin{align}
h_{1}(z;\phi_{*}^{(1)}) & =H^{-3}(\phi_{*}^{(1)})e^{-zH^{-3}(\phi_{*}^{(1)})},\quad h_{1}(z;\phi_{*}^{(2)})=0;\label{eq:h1 bc for 12}\\
h_{2}(z;\phi_{*}^{(1)}) & =0,\quad h_{2}(z;\phi_{*}^{(2)})=H^{-3}(\phi_{*}^{(2)})e^{-zH^{-3}(\phi_{*}^{(2)})}.\label{eq:h2 bc for 12}\end{align}
In the present toy model the diffusion coefficient is set to zero
at reheating, so the formalism developed in Sec.~\ref{sub:FPRV-distribution-of-Q}
needs to be modified. We will first solve Eq.~(\ref{eq:h equ toy})
analytically in the no-diffusion intervals of $\phi$ and obtain the
boundary conditions for $h$ at the boundaries of the self-reproduction
regime $\left[\phi_{1},\phi_{2}\right]$ where $D(\phi)\neq0$. Then
the methods of Sec.~\ref{sub:FPRV-distribution-of-Q} will be applicable
to the boundary value problem for the interval $\left[\phi_{1},\phi_{2}\right]$.

Implementing this idea in the first no-diffusion region $\phi_{*}^{(1)}<\phi<\phi_{1}$,
we use the solution~(\ref{eq:g sol nodiff 12}) for $g(z;\phi)$
and reduce Eq.~(\ref{eq:h equ toy}) to\begin{equation}
\partial_{\phi}h_{j}+\frac{3H}{v}\left[1-zH^{-3}(\phi_{*}^{(1)})\exp\!\left(-3\!\int_{\phi_{*}^{(1)}}^{\phi}\!\!\frac{H}{v}d\phi\right)\!\right]h_{j}=0.\end{equation}
This equation is easily integrated together with the boundary conditions~(\ref{eq:h1 bc for 12})--(\ref{eq:h2 bc for 12})
and yields the values of $h_{j}$ at $\phi_{1}$,

\begin{equation}
h_{j}(z;\phi_{1})=\delta_{j1}H^{-3}(\phi_{*}^{(1)})\exp\left[3N_{1}-zH^{-3}(\phi_{*}^{(1)})e^{3N_{1}}\right].\end{equation}
Similarly we can determine the values $h_{j}(z;\phi_{2})$. Since
the value $z=z_{*}$ is important for the present calculation, we
now find the values of $h_{j}$ at $z=z_{*}$ using the assumption
$N_{2}\gg N_{1}$ and the estimate~(\ref{eq:z star estimate}),\begin{equation}
h_{j}(z_{*};\phi_{i})\approx\delta_{ij}\frac{\exp\left[3N_{i}+\frac{1}{2}\delta_{i2}\right]}{H^{3}(\phi_{*}^{(i)})},\quad i,j=1,2.\label{eq:h bc new}\end{equation}
We have thus reduced the problem of determining $h_{j}(z;\phi)$ to
the boundary-value problem for the interval $\left[\phi_{1},\phi_{2}\right]$
where the methods of Sec.~\ref{sub:FPRV-distribution-of-Q} apply
but the boundary conditions are given by Eq.~(\ref{eq:h bc new}). 

The next step, according to Sec.~\ref{sub:FPRV-distribution-of-Q},
is to compute the eigenfunction $f_{0}(z_{*};\phi)$ of the operator
\begin{equation}
\hat{\tilde{L}}\equiv\frac{D_{0}}{H_{0}}\partial_{\phi}\partial_{\phi}+3(\ln g(z_{*};\phi)+1)\end{equation}
such that\begin{equation}
\hat{\tilde{L}}f_{0}=0;\quad f_{0}(z_{*};\phi_{1,2})=0.\end{equation}
As we have shown, this eigenfunction with eigenvalue 0 exists precisely
at $z=z_{*}$. Once this eigenfunction is computed, the ratio of the
RV-regulated mean volumes in channels 1 and 2 will be expressed through
the derivatives of $f_{0}$ at the endpoints and through the modified
boundary conditions~(\ref{eq:h bc new}) as follows,\begin{equation}
\frac{P(2)}{P(1)}=\frac{h_{2}(z_{*};\phi_{2})}{h_{1}(z_{*};\phi_{1})}\frac{\left|\partial_{\phi}f_{0}(z_{*};\phi_{2})\right|}{\partial_{\phi}f_{0}(z_{*};\phi_{1})}.\label{eq:P2P1 pre}\end{equation}
The absolute value is taken to compensate for the negative sign of
the derivative $\partial_{\phi}f_{0}$ at the right boundary point
(assuming that $f_{0}\geq0$ everywhere). Since $h_{1,2}(z_{*};\phi_{1,2})$
are already known, it remains to derive an estimate for $f_{0}(z_{*};\phi)$.

\begin{figure}
\begin{centering}\psfrag{V}{$\tilde V (\phi)$} \psfrag{f1}{$\phi_1$} \psfrag{f9}{$\phi_0$} \psfrag{f2}{$\phi_2$} \psfrag{f0}{$f_0(z_{*};\phi)$} \psfrag{f}{$\phi$} \includegraphics[width=0.4\textwidth]{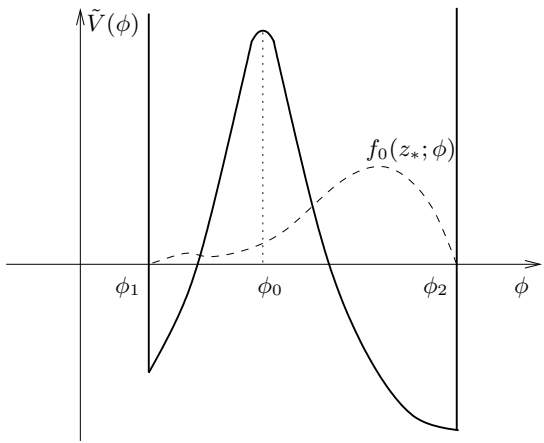}\par\end{centering}

\caption{Sketch of the eigenfunction $f_{0}(z_{*};\phi)$ (dashed line) interpreted
as the wavefunction of a stationary state with zero energy in the
potential $\tilde{V}(\phi)$ (solid line). Due to exponential suppression
by the potential barrier, the amplitude of $f_{0}$ in the right region
is exponentially larger than that in the left region. \label{fig:Wavefunction-potential}}
\end{figure}

The eigenvalue equation $\hat{\tilde{L}}f_{0}=0$ formally resembles
a one-dimensional Schrödinger equation with the coordinate $\phi$
and the {}``potential'' \begin{equation}
\tilde{V}(\phi)\equiv-\frac{12\pi^{2}}{H_{0}^{2}}\left(\ln g(z_{*};\phi)+1\right).\label{eq:V tilde def}\end{equation}
The eigenfunction $f_{0}(z_{*};\phi)$ is then interpreted as the
{}``wavefunction'' of a stationary state with zero energy and zero
boundary conditions at $\phi=\phi_{1,2}$. According to Eqs.~(\ref{eq:g approx 2})
and (\ref{eq:g approx 1}), the function $\tilde{V}(\phi)$ has a
maximum at $\phi\approx\phi_{0}$ (see Fig.~\ref{fig:Wavefunction-potential}),
while its values at the endpoints are\begin{equation}
\tilde{V}(\phi_{1})\approx-\frac{12\pi^{2}}{H_{0}^{2}},\quad\tilde{V}(\phi_{2})\approx-\frac{18\pi^{2}}{H_{0}^{2}}.\end{equation}
Using the terminology of quantum mechanics, there is a potential barrier
separating two classically allowed regions near $\phi=\phi_{1}$ and
$\phi=\phi_{2}$. Since the {}``potential well'' at $\phi=\phi_{2}$
is deeper, the ground state is approximately the ground state of that
one well, with an exponentially small amplitude of being near $\phi=\phi_{1}$.
The shape of the eigenfunction is sketched in Fig.~\ref{fig:Wavefunction-potential}.
The exponential suppression of the amplitude near $\phi=\phi_{1}$
can be found using the WKB approximation, which yields\begin{equation}
\frac{\left|\partial_{\phi}f_{0}(z_{*};\phi_{2})\right|}{\partial_{\phi}f_{0}(z_{*};\phi_{1})}=A_{21}\exp\left[\int_{\tilde{\phi}_{1}}^{\tilde{\phi}_{2}}\sqrt{\tilde{V}(\phi)}d\phi\right],\end{equation}
where $\tilde{\phi}_{1,2}$ are the turning points such that $\tilde{V}(\tilde{\phi}_{1,2})=0$.
The pre-exponential factor $A_{21}$ is of order 1 and can, in principle,
be obtained from a more detailed matching of the WKB-approximated
solution across the barrier to the solutions in the {}``classically
allowed'' regions, or by determining the solution $f_{0}(z_{*};\phi)$
numerically. However, we will omit this calculation since the main
result will consist of an exponentially large factor. That factor
can be estimated using Eqs.~(\ref{eq:g approx 2}), (\ref{eq:g approx 1}),
and (\ref{eq:V tilde def}) as\begin{equation}
\int_{\tilde{\phi}_{1}}^{\tilde{\phi}_{2}}\sqrt{\tilde{V}(\phi)}d\phi\approx2\int_{\phi_{1}}^{\phi_{0}}\sqrt{-\frac{12\pi^{2}}{H_{0}^{2}}\ln g}\, d\phi\approx\frac{3\pi^{2}}{\sqrt{2}H_{0}^{2}}\left(\phi_{2}-\phi_{1}\right)^{2}.\end{equation}
Hence, the ratio~(\ref{eq:P2P1 pre}) is simplified to\begin{equation}
\frac{P(2)}{P(1)}=A_{21}\frac{H^{-3}(\phi_{*}^{(2)})}{H^{-3}(\phi_{*}^{(1)})}\frac{e^{3N_{2}+\frac{1}{2}}}{e^{3N_{1}}}\exp\left[\frac{3\pi^{2}}{\sqrt{2}H_{0}^{2}}\left(\phi_{2}-\phi_{1}\right)^{2}\right].\end{equation}
This is the main result quoted above in Eq.~(\ref{eq:RV result toy}).

\section*{Acknowledgments}

The author thanks Martin Bucher, Jaume Garriga, Andrei Linde, Vitaly
Vanchurin, Takahiro Tanaka, and Alex Vilenkin for valuable discussions.
The author gratefully acknowledges the hospitality of the Yukawa Institute
of Theoretical Physics (University of Kyoto) where part of this work
was completed. The stay of the author at the YITP was supported by
the Yukawa International Program for Quark-Hadron Sciences.

\bibliographystyle{myphysrev}
\bibliography{EI2}

\begin{thebibliography}{61}
\expandafter\ifx\csname natexlab\endcsname\relax\def\natexlab#1{#1}\fi
\expandafter\ifx\csname bibnamefont\endcsname\relax
  \def\bibnamefont#1{#1}\fi
\expandafter\ifx\csname bibfnamefont\endcsname\relax
  \def\bibfnamefont#1{#1}\fi
\expandafter\ifx\csname citenamefont\endcsname\relax
  \def\citenamefont#1{#1}\fi
\expandafter\ifx\csname url\endcsname\relax
  \def\url#1{\texttt{#1}}\fi
\expandafter\ifx\csname urlprefix\endcsname\relax\def\urlprefix{URL }\fi
\providecommand{\bibinfo}[2]{#2}
\providecommand{\eprint}[2][]{\url{#2}}

\bibitem[{\citenamefont{Bousso and Polchinski}(2000)}]{Bousso:2000xa}
\bibinfo{author}{\bibfnamefont{R.}~\bibnamefont{Bousso}} \bibnamefont{and}
  \bibinfo{author}{\bibfnamefont{J.}~\bibnamefont{Polchinski}},
  \emph{\bibinfo{title}{Quantization of four-form fluxes and dynamical
  neutralization of the cosmological constant}}, \bibinfo{journal}{JHEP}
  \textbf{\bibinfo{volume}{06}}, \bibinfo{pages}{006} (\bibinfo{year}{2000}),
  \eprint{hep-th/0004134}.

\bibitem[{\citenamefont{Susskind}(2003)}]{Susskind:2003kw}
\bibinfo{author}{\bibfnamefont{L.}~\bibnamefont{Susskind}},
  \emph{\bibinfo{title}{The anthropic landscape of string theory}}
  (\bibinfo{year}{2003}), \eprint{hep-th/0302219}.

\bibitem[{\citenamefont{Douglas}(2003)}]{Douglas:2003um}
\bibinfo{author}{\bibfnamefont{M.~R.} \bibnamefont{Douglas}},
  \emph{\bibinfo{title}{{The statistics of string / M theory vacua}}},
  \bibinfo{journal}{JHEP} \textbf{\bibinfo{volume}{05}}, \bibinfo{pages}{046}
  (\bibinfo{year}{2003}), \eprint{hep-th/0303194}.

\bibitem[{\citenamefont{Garriga and Vilenkin}(1998)}]{Garriga:1997ef}
\bibinfo{author}{\bibfnamefont{J.}~\bibnamefont{Garriga}} \bibnamefont{and}
  \bibinfo{author}{\bibfnamefont{A.}~\bibnamefont{Vilenkin}},
  \emph{\bibinfo{title}{Recycling universe}}, \bibinfo{journal}{Phys. Rev.}
  \textbf{\bibinfo{volume}{D57}}, \bibinfo{pages}{2230} (\bibinfo{year}{1998}),
  \eprint{astro-ph/9707292}.

\bibitem[{\citenamefont{Garcia-Bellido
  et~al.}(1994)\citenamefont{Garcia-Bellido, Linde, and
  Linde}}]{Garcia-Bellido:1993wn}
\bibinfo{author}{\bibfnamefont{J.}~\bibnamefont{Garcia-Bellido}},
  \bibinfo{author}{\bibfnamefont{A.~D.} \bibnamefont{Linde}}, \bibnamefont{and}
  \bibinfo{author}{\bibfnamefont{D.~A.} \bibnamefont{Linde}},
  \emph{\bibinfo{title}{Fluctuations of the gravitational constant in the
  inflationary brans-dicke cosmology}}, \bibinfo{journal}{Phys. Rev.}
  \textbf{\bibinfo{volume}{D50}}, \bibinfo{pages}{730} (\bibinfo{year}{1994}),
  \eprint{astro-ph/9312039}.

\bibitem[{\citenamefont{Garcia-Bellido}(1994)}]{Garcia-Bellido:1994vz}
\bibinfo{author}{\bibfnamefont{J.}~\bibnamefont{Garcia-Bellido}},
  \emph{\bibinfo{title}{Jordan-brans-dicke stochastic inflation}},
  \bibinfo{journal}{Nucl. Phys.} \textbf{\bibinfo{volume}{B423}},
  \bibinfo{pages}{221} (\bibinfo{year}{1994}), \eprint{astro-ph/9401042}.

\bibitem[{\citenamefont{Linde et~al.}(1994)\citenamefont{Linde, Linde, and
  Mezhlumian}}]{Linde:1993xx}
\bibinfo{author}{\bibfnamefont{A.~D.} \bibnamefont{Linde}},
  \bibinfo{author}{\bibfnamefont{D.~A.} \bibnamefont{Linde}}, \bibnamefont{and}
  \bibinfo{author}{\bibfnamefont{A.}~\bibnamefont{Mezhlumian}},
  \emph{\bibinfo{title}{{From the Big Bang theory to the theory of a stationary
  universe}}}, \bibinfo{journal}{Phys. Rev.} \textbf{\bibinfo{volume}{D49}},
  \bibinfo{pages}{1783} (\bibinfo{year}{1994}), \eprint{gr-qc/9306035}.

\bibitem[{\citenamefont{Guth}(2000)}]{Guth:2000ka}
\bibinfo{author}{\bibfnamefont{A.~H.} \bibnamefont{Guth}},
  \emph{\bibinfo{title}{Inflation and eternal inflation}},
  \bibinfo{journal}{Phys. Rept.} \textbf{\bibinfo{volume}{333}},
  \bibinfo{pages}{555} (\bibinfo{year}{2000}), \eprint{astro-ph/0002156}.

\bibitem[{\citenamefont{Winitzki}(2008{\natexlab{a}})}]{Winitzki:2006rn}
\bibinfo{author}{\bibfnamefont{S.}~\bibnamefont{Winitzki}},
  \emph{\bibinfo{title}{Predictions in eternal inflation}},
  \bibinfo{journal}{Lect. Notes Phys.} \textbf{\bibinfo{volume}{738}},
  \bibinfo{pages}{157} (\bibinfo{year}{2008}{\natexlab{a}}),
  \eprint{gr-qc/0612164}.

\bibitem[{\citenamefont{Garcia-Bellido and
  Linde}(1995)}]{Garcia-Bellido:1994ci}
\bibinfo{author}{\bibfnamefont{J.}~\bibnamefont{Garcia-Bellido}}
  \bibnamefont{and} \bibinfo{author}{\bibfnamefont{A.~D.} \bibnamefont{Linde}},
  \emph{\bibinfo{title}{Stationarity of inflation and predictions of quantum
  cosmology}}, \bibinfo{journal}{Phys. Rev.} \textbf{\bibinfo{volume}{D51}},
  \bibinfo{pages}{429} (\bibinfo{year}{1995}), \eprint{hep-th/9408023}.

\bibitem[{\citenamefont{Vilenkin}(1995{\natexlab{a}})}]{Vilenkin:1994ua}
\bibinfo{author}{\bibfnamefont{A.}~\bibnamefont{Vilenkin}},
  \emph{\bibinfo{title}{Predictions from quantum cosmology}},
  \bibinfo{journal}{Phys. Rev. Lett.} \textbf{\bibinfo{volume}{74}},
  \bibinfo{pages}{846} (\bibinfo{year}{1995}{\natexlab{a}}),
  \eprint{gr-qc/9406010}.

\bibitem[{\citenamefont{Vilenkin}(1995{\natexlab{b}})}]{Vilenkin:1995yd}
\bibinfo{author}{\bibfnamefont{A.}~\bibnamefont{Vilenkin}},
  \emph{\bibinfo{title}{Making predictions in eternally inflating universe}},
  \bibinfo{journal}{Phys. Rev.} \textbf{\bibinfo{volume}{D52}},
  \bibinfo{pages}{3365} (\bibinfo{year}{1995}{\natexlab{b}}),
  \eprint{gr-qc/9505031}.

\bibitem[{\citenamefont{Borde and Vilenkin}(1994)}]{Borde:1993xh}
\bibinfo{author}{\bibfnamefont{A.}~\bibnamefont{Borde}} \bibnamefont{and}
  \bibinfo{author}{\bibfnamefont{A.}~\bibnamefont{Vilenkin}},
  \emph{\bibinfo{title}{Eternal inflation and the initial singularity}},
  \bibinfo{journal}{Phys. Rev. Lett.} \textbf{\bibinfo{volume}{72}},
  \bibinfo{pages}{3305} (\bibinfo{year}{1994}), \eprint{gr-qc/9312022}.

\bibitem[{\citenamefont{Creminelli et~al.}(2008)\citenamefont{Creminelli,
  Dubovsky, Nicolis, Senatore, and Zaldarriaga}}]{Creminelli:2008es}
\bibinfo{author}{\bibfnamefont{P.}~\bibnamefont{Creminelli}},
  \bibinfo{author}{\bibfnamefont{S.}~\bibnamefont{Dubovsky}},
  \bibinfo{author}{\bibfnamefont{A.}~\bibnamefont{Nicolis}},
  \bibinfo{author}{\bibfnamefont{L.}~\bibnamefont{Senatore}}, \bibnamefont{and}
  \bibinfo{author}{\bibfnamefont{M.}~\bibnamefont{Zaldarriaga}},
  \emph{\bibinfo{title}{The phase transition to slow-roll eternal inflation}}
  (\bibinfo{year}{2008}), \eprint{arXiv:0802.1067 [hep-th]}.

\bibitem[{\citenamefont{Winitzki}(2002)}]{Winitzki:2001np}
\bibinfo{author}{\bibfnamefont{S.}~\bibnamefont{Winitzki}},
  \emph{\bibinfo{title}{The eternal fractal in the universe}},
  \bibinfo{journal}{Phys. Rev.} \textbf{\bibinfo{volume}{D65}},
  \bibinfo{pages}{083506} (\bibinfo{year}{2002}), \eprint{gr-qc/0111048}.

\bibitem[{\citenamefont{Winitzki}(2005{\natexlab{a}})}]{Winitzki:2005fy}
\bibinfo{author}{\bibfnamefont{S.}~\bibnamefont{Winitzki}},
  \emph{\bibinfo{title}{Drawing conformal diagrams for a fractal landscape}},
  \bibinfo{journal}{Phys. Rev.} \textbf{\bibinfo{volume}{D71}},
  \bibinfo{pages}{123523} (\bibinfo{year}{2005}{\natexlab{a}}),
  \eprint{gr-qc/0503061}.

\bibitem[{\citenamefont{Vanchurin}(2007)}]{Vanchurin:2006xp}
\bibinfo{author}{\bibfnamefont{V.}~\bibnamefont{Vanchurin}},
  \emph{\bibinfo{title}{{Geodesic measures of the landscape}}},
  \bibinfo{journal}{Phys. Rev.} \textbf{\bibinfo{volume}{D75}},
  \bibinfo{pages}{023524} (\bibinfo{year}{2007}), \eprint{hep-th/0612215}.

\bibitem[{\citenamefont{Aguirre et~al.}(2006)\citenamefont{Aguirre, Gratton,
  and Johnson}}]{Aguirre:2006ak}
\bibinfo{author}{\bibfnamefont{A.}~\bibnamefont{Aguirre}},
  \bibinfo{author}{\bibfnamefont{S.}~\bibnamefont{Gratton}}, \bibnamefont{and}
  \bibinfo{author}{\bibfnamefont{M.~C.} \bibnamefont{Johnson}},
  \emph{\bibinfo{title}{Hurdles for recent measures in eternal inflation}}
  (\bibinfo{year}{2006}), \eprint{hep-th/0611221}.

\bibitem[{\citenamefont{Vilenkin}(2007{\natexlab{a}})}]{Vilenkin:2006xv}
\bibinfo{author}{\bibfnamefont{A.}~\bibnamefont{Vilenkin}},
  \emph{\bibinfo{title}{A measure of the multiverse}}, \bibinfo{journal}{J.
  Phys.} \textbf{\bibinfo{volume}{A40}}, \bibinfo{pages}{6777}
  (\bibinfo{year}{2007}{\natexlab{a}}), \eprint{hep-th/0609193}.

\bibitem[{\citenamefont{Guth}(2007)}]{Guth:2007ng}
\bibinfo{author}{\bibfnamefont{A.~H.} \bibnamefont{Guth}},
  \emph{\bibinfo{title}{Eternal inflation and its implications}},
  \bibinfo{journal}{J. Phys.} \textbf{\bibinfo{volume}{A40}},
  \bibinfo{pages}{6811} (\bibinfo{year}{2007}), \eprint{hep-th/0702178}.

\bibitem[{\citenamefont{Linde}(2007{\natexlab{a}})}]{Linde:2007nm}
\bibinfo{author}{\bibfnamefont{A.}~\bibnamefont{Linde}},
  \emph{\bibinfo{title}{Towards a gauge invariant volume-weighted probability
  measure for eternal inflation}}, \bibinfo{journal}{JCAP}
  \textbf{\bibinfo{volume}{0706}}, \bibinfo{pages}{017}
  (\bibinfo{year}{2007}{\natexlab{a}}), \eprint{arXiv:0705.1160 [hep-th]}.

\bibitem[{\citenamefont{Vanchurin et~al.}(2000)\citenamefont{Vanchurin,
  Vilenkin, and Winitzki}}]{Vanchurin:1999iv}
\bibinfo{author}{\bibfnamefont{V.}~\bibnamefont{Vanchurin}},
  \bibinfo{author}{\bibfnamefont{A.}~\bibnamefont{Vilenkin}}, \bibnamefont{and}
  \bibinfo{author}{\bibfnamefont{S.}~\bibnamefont{Winitzki}},
  \emph{\bibinfo{title}{Predictability crisis in inflationary cosmology and its
  resolution}}, \bibinfo{journal}{Phys. Rev.} \textbf{\bibinfo{volume}{D61}},
  \bibinfo{pages}{083507} (\bibinfo{year}{2000}), \eprint{gr-qc/9905097}.

\bibitem[{\citenamefont{Winitzki}(2005{\natexlab{b}})}]{Winitzki:2005ya}
\bibinfo{author}{\bibfnamefont{S.}~\bibnamefont{Winitzki}},
  \emph{\bibinfo{title}{On time-reparametrization invariance in eternal
  inflation}}, \bibinfo{journal}{Phys. Rev.} \textbf{\bibinfo{volume}{D71}},
  \bibinfo{pages}{123507} (\bibinfo{year}{2005}{\natexlab{b}}),
  \eprint{gr-qc/0504084}.

\bibitem[{\citenamefont{Winitzki and Vilenkin}(1996)}]{Winitzki:1995pg}
\bibinfo{author}{\bibfnamefont{S.}~\bibnamefont{Winitzki}} \bibnamefont{and}
  \bibinfo{author}{\bibfnamefont{A.}~\bibnamefont{Vilenkin}},
  \emph{\bibinfo{title}{Uncertainties of predictions in models of eternal
  inflation}}, \bibinfo{journal}{Phys. Rev.} \textbf{\bibinfo{volume}{D53}},
  \bibinfo{pages}{4298} (\bibinfo{year}{1996}), \eprint{gr-qc/9510054}.

\bibitem[{\citenamefont{Linde and Mezhlumian}(1996)}]{Linde:1995uf}
\bibinfo{author}{\bibfnamefont{A.~D.} \bibnamefont{Linde}} \bibnamefont{and}
  \bibinfo{author}{\bibfnamefont{A.}~\bibnamefont{Mezhlumian}},
  \emph{\bibinfo{title}{On regularization scheme dependence of predictions in
  inflationary cosmology}}, \bibinfo{journal}{Phys. Rev.}
  \textbf{\bibinfo{volume}{D53}}, \bibinfo{pages}{4267} (\bibinfo{year}{1996}),
  \eprint{gr-qc/9511058}.

\bibitem[{\citenamefont{Linde et~al.}(1995)\citenamefont{Linde, Linde, and
  Mezhlumian}}]{Linde:1994gy}
\bibinfo{author}{\bibfnamefont{A.~D.} \bibnamefont{Linde}},
  \bibinfo{author}{\bibfnamefont{D.~A.} \bibnamefont{Linde}}, \bibnamefont{and}
  \bibinfo{author}{\bibfnamefont{A.}~\bibnamefont{Mezhlumian}},
  \emph{\bibinfo{title}{Do we live in the center of the world?}},
  \bibinfo{journal}{Phys. Lett.} \textbf{\bibinfo{volume}{B345}},
  \bibinfo{pages}{203} (\bibinfo{year}{1995}), \eprint{hep-th/9411111}.

\bibitem[{\citenamefont{Vilenkin}(1998)}]{Vilenkin:1998kr}
\bibinfo{author}{\bibfnamefont{A.}~\bibnamefont{Vilenkin}},
  \emph{\bibinfo{title}{Unambiguous probabilities in an eternally inflating
  universe}}, \bibinfo{journal}{Phys. Rev. Lett.}
  \textbf{\bibinfo{volume}{81}}, \bibinfo{pages}{5501} (\bibinfo{year}{1998}),
  \eprint{hep-th/9806185}.

\bibitem[{\citenamefont{Dyson et~al.}(2002)\citenamefont{Dyson, Kleban, and
  Susskind}}]{Dyson:2002pf}
\bibinfo{author}{\bibfnamefont{L.}~\bibnamefont{Dyson}},
  \bibinfo{author}{\bibfnamefont{M.}~\bibnamefont{Kleban}}, \bibnamefont{and}
  \bibinfo{author}{\bibfnamefont{L.}~\bibnamefont{Susskind}},
  \emph{\bibinfo{title}{{Disturbing implications of a cosmological constant}}},
  \bibinfo{journal}{JHEP} \textbf{\bibinfo{volume}{10}}, \bibinfo{pages}{011}
  (\bibinfo{year}{2002}), \eprint{hep-th/0208013}.

\bibitem[{\citenamefont{Albrecht and Sorbo}(2004)}]{Albrecht:2004ke}
\bibinfo{author}{\bibfnamefont{A.}~\bibnamefont{Albrecht}} \bibnamefont{and}
  \bibinfo{author}{\bibfnamefont{L.}~\bibnamefont{Sorbo}},
  \emph{\bibinfo{title}{Can the universe afford inflation?}},
  \bibinfo{journal}{Phys. Rev.} \textbf{\bibinfo{volume}{D70}},
  \bibinfo{pages}{063528} (\bibinfo{year}{2004}), \eprint{hep-th/0405270}.

\bibitem[{\citenamefont{Page}(2006{\natexlab{a}})}]{Page:2006dt}
\bibinfo{author}{\bibfnamefont{D.~N.} \bibnamefont{Page}},
  \emph{\bibinfo{title}{{Is our universe likely to decay within 20 billion
  years?}}} (\bibinfo{year}{2006}{\natexlab{a}}), \eprint{hep-th/0610079}.

\bibitem[{\citenamefont{Linde}(2007{\natexlab{b}})}]{Linde:2006nw}
\bibinfo{author}{\bibfnamefont{A.}~\bibnamefont{Linde}},
  \emph{\bibinfo{title}{{Sinks in the landscape, Boltzmann brains, and the
  cosmological constant oroblem}}}, \bibinfo{journal}{JCAP}
  \textbf{\bibinfo{volume}{0701}}, \bibinfo{pages}{022}
  (\bibinfo{year}{2007}{\natexlab{b}}), \eprint{hep-th/0611043}.

\bibitem[{\citenamefont{Vilenkin}(2007{\natexlab{b}})}]{Vilenkin:2006qg}
\bibinfo{author}{\bibfnamefont{A.}~\bibnamefont{Vilenkin}},
  \emph{\bibinfo{title}{{Freak observers and the measure of the multiverse}}},
  \bibinfo{journal}{JHEP} \textbf{\bibinfo{volume}{01}}, \bibinfo{pages}{092}
  (\bibinfo{year}{2007}{\natexlab{b}}), \eprint{hep-th/0611271}.

\bibitem[{\citenamefont{Page}(2006{\natexlab{b}})}]{Page:2006ys}
\bibinfo{author}{\bibfnamefont{D.~N.} \bibnamefont{Page}},
  \emph{\bibinfo{title}{{Return of the Boltzmann brains}}}
  (\bibinfo{year}{2006}{\natexlab{b}}), \eprint{hep-th/0611158}.

\bibitem[{\citenamefont{Bousso et~al.}(2007)\citenamefont{Bousso, Freivogel,
  and Yang}}]{Bousso:2007nd}
\bibinfo{author}{\bibfnamefont{R.}~\bibnamefont{Bousso}},
  \bibinfo{author}{\bibfnamefont{B.}~\bibnamefont{Freivogel}},
  \bibnamefont{and} \bibinfo{author}{\bibfnamefont{I.-S.} \bibnamefont{Yang}},
  \emph{\bibinfo{title}{{Boltzmann babies in the proper time measure}}}
  (\bibinfo{year}{2007}), \eprint{arXiv:0712.3324 [hep-th]}.

\bibitem[{\citenamefont{Gott~III}(2008)}]{Gott:2008ii}
\bibinfo{author}{\bibfnamefont{J.~R.} \bibnamefont{Gott~III}},
  \emph{\bibinfo{title}{{Boltzmann brains--I'd rather see than be one}}}
  (\bibinfo{year}{2008}), \eprint{arXiv:0802.0233 [gr-qc]}.

\bibitem[{\citenamefont{Bousso}(2006)}]{Bousso:2006ev}
\bibinfo{author}{\bibfnamefont{R.}~\bibnamefont{Bousso}},
  \emph{\bibinfo{title}{Holographic probabilities in eternal inflation}},
  \bibinfo{journal}{Phys. Rev. Lett.} \textbf{\bibinfo{volume}{97}},
  \bibinfo{pages}{191302} (\bibinfo{year}{2006}), \eprint{hep-th/0605263}.

\bibitem[{\citenamefont{Bousso et~al.}(2006)\citenamefont{Bousso, Freivogel,
  and Yang}}]{Bousso:2006ge}
\bibinfo{author}{\bibfnamefont{R.}~\bibnamefont{Bousso}},
  \bibinfo{author}{\bibfnamefont{B.}~\bibnamefont{Freivogel}},
  \bibnamefont{and} \bibinfo{author}{\bibfnamefont{I.-S.} \bibnamefont{Yang}},
  \emph{\bibinfo{title}{Eternal inflation: The inside story}}
  (\bibinfo{year}{2006}), \eprint{hep-th/0606114}.

\bibitem[{\citenamefont{Tolley and Wyman}(2008)}]{Tolley:2008na}
\bibinfo{author}{\bibfnamefont{A.~J.} \bibnamefont{Tolley}} \bibnamefont{and}
  \bibinfo{author}{\bibfnamefont{M.}~\bibnamefont{Wyman}},
  \emph{\bibinfo{title}{{Stochastic inflation revisited: non-slow roll
  statistics and DBI inflation}}}, \bibinfo{journal}{JCAP}
  \textbf{\bibinfo{volume}{0804}}, \bibinfo{pages}{028} (\bibinfo{year}{2008}),
  \eprint{0801.1854}.

\bibitem[{\citenamefont{Pogosian and Vilenkin}(2006)}]{Pogosian:2006fx}
\bibinfo{author}{\bibfnamefont{L.}~\bibnamefont{Pogosian}} \bibnamefont{and}
  \bibinfo{author}{\bibfnamefont{A.}~\bibnamefont{Vilenkin}},
  \emph{\bibinfo{title}{Anthropic predictions for vacuum energy and neutrino
  masses in the light of wmap-3}} (\bibinfo{year}{2006}),
  \eprint{astro-ph/0611573}.

\bibitem[{\citenamefont{Maor et~al.}(2007)\citenamefont{Maor, Krauss, and
  Starkman}}]{Maor:2007wq}
\bibinfo{author}{\bibfnamefont{I.}~\bibnamefont{Maor}},
  \bibinfo{author}{\bibfnamefont{L.}~\bibnamefont{Krauss}}, \bibnamefont{and}
  \bibinfo{author}{\bibfnamefont{G.}~\bibnamefont{Starkman}},
  \emph{\bibinfo{title}{Anthropics and myopics: Conditional probabilities and
  the cosmological constant}} (\bibinfo{year}{2007}), \eprint{arXiv:0709.0502
  [hep-th]}.

\bibitem[{\citenamefont{Garriga and Vilenkin}(2007)}]{Garriga:2007wz}
\bibinfo{author}{\bibfnamefont{J.}~\bibnamefont{Garriga}} \bibnamefont{and}
  \bibinfo{author}{\bibfnamefont{A.}~\bibnamefont{Vilenkin}},
  \emph{\bibinfo{title}{Prediction and explanation in the multiverse}}
  (\bibinfo{year}{2007}), \eprint{arXiv:0711.2559 [hep-th]}.

\bibitem[{\citenamefont{Hartle and Srednicki}(2007)}]{Hartle:2007zv}
\bibinfo{author}{\bibfnamefont{J.~B.} \bibnamefont{Hartle}} \bibnamefont{and}
  \bibinfo{author}{\bibfnamefont{M.}~\bibnamefont{Srednicki}},
  \emph{\bibinfo{title}{Are we typical?}}, \bibinfo{journal}{Phys. Rev.}
  \textbf{\bibinfo{volume}{D75}}, \bibinfo{pages}{123523}
  (\bibinfo{year}{2007}), \eprint{arXiv:0704.2630 [hep-th]}.

\bibitem[{\citenamefont{Gratton and Turok}(2005)}]{Gratton:2005bi}
\bibinfo{author}{\bibfnamefont{S.}~\bibnamefont{Gratton}} \bibnamefont{and}
  \bibinfo{author}{\bibfnamefont{N.}~\bibnamefont{Turok}},
  \emph{\bibinfo{title}{Langevin analysis of eternal inflation}},
  \bibinfo{journal}{Phys. Rev.} \textbf{\bibinfo{volume}{D72}},
  \bibinfo{pages}{043507} (\bibinfo{year}{2005}), \eprint{hep-th/0503063}.

\bibitem[{\citenamefont{Garriga et~al.}(2006)\citenamefont{Garriga,
  Schwartz-Perlov, Vilenkin, and Winitzki}}]{Garriga:2005av}
\bibinfo{author}{\bibfnamefont{J.}~\bibnamefont{Garriga}},
  \bibinfo{author}{\bibfnamefont{D.}~\bibnamefont{Schwartz-Perlov}},
  \bibinfo{author}{\bibfnamefont{A.}~\bibnamefont{Vilenkin}}, \bibnamefont{and}
  \bibinfo{author}{\bibfnamefont{S.}~\bibnamefont{Winitzki}},
  \emph{\bibinfo{title}{Probabilities in the inflationary multiverse}},
  \bibinfo{journal}{JCAP} \textbf{\bibinfo{volume}{0601}}, \bibinfo{pages}{017}
  (\bibinfo{year}{2006}), \eprint{hep-th/0509184}.

\bibitem[{\citenamefont{Easther et~al.}(2006)\citenamefont{Easther, Lim, and
  Martin}}]{Easther:2005wi}
\bibinfo{author}{\bibfnamefont{R.}~\bibnamefont{Easther}},
  \bibinfo{author}{\bibfnamefont{E.~A.} \bibnamefont{Lim}}, \bibnamefont{and}
  \bibinfo{author}{\bibfnamefont{M.~R.} \bibnamefont{Martin}},
  \emph{\bibinfo{title}{Counting pockets with world lines in eternal
  inflation}}, \bibinfo{journal}{JCAP} \textbf{\bibinfo{volume}{0603}},
  \bibinfo{pages}{016} (\bibinfo{year}{2006}), \eprint{astro-ph/0511233}.

\bibitem[{\citenamefont{Vanchurin and Vilenkin}(2006)}]{Vanchurin:2006qp}
\bibinfo{author}{\bibfnamefont{V.}~\bibnamefont{Vanchurin}} \bibnamefont{and}
  \bibinfo{author}{\bibfnamefont{A.}~\bibnamefont{Vilenkin}},
  \emph{\bibinfo{title}{Eternal observers and bubble abundances in the
  landscape}}, \bibinfo{journal}{Phys. Rev.} \textbf{\bibinfo{volume}{D74}},
  \bibinfo{pages}{043520} (\bibinfo{year}{2006}), \eprint{hep-th/0605015}.

\bibitem[{\citenamefont{Clifton et~al.}(2007)\citenamefont{Clifton, Shenker,
  and Sivanandam}}]{Clifton:2007bn}
\bibinfo{author}{\bibfnamefont{T.}~\bibnamefont{Clifton}},
  \bibinfo{author}{\bibfnamefont{S.}~\bibnamefont{Shenker}}, \bibnamefont{and}
  \bibinfo{author}{\bibfnamefont{N.}~\bibnamefont{Sivanandam}},
  \emph{\bibinfo{title}{{Volume Weighted Measures of Eternal Inflation in the
  Bousso-Polchinski Landscape}}}, \bibinfo{journal}{JHEP}
  \textbf{\bibinfo{volume}{09}}, \bibinfo{pages}{034} (\bibinfo{year}{2007}),
  \eprint{arXiv:0706.3201 [hep-th]}.

\bibitem[{\citenamefont{Hawking}(2007{\natexlab{a}})}]{Hawking:2007zz}
\bibinfo{author}{\bibfnamefont{S.~W.} \bibnamefont{Hawking}},
  \emph{\bibinfo{title}{{The measure of the universe}}}, \bibinfo{journal}{AIP
  Conf. Proc.} \textbf{\bibinfo{volume}{957}}, \bibinfo{pages}{79}
  (\bibinfo{year}{2007}{\natexlab{a}}).

\bibitem[{\citenamefont{Hartle et~al.}(2007)\citenamefont{Hartle, Hawking, and
  Hertog}}]{Hartle:2007gi}
\bibinfo{author}{\bibfnamefont{J.~B.} \bibnamefont{Hartle}},
  \bibinfo{author}{\bibfnamefont{S.~W.} \bibnamefont{Hawking}},
  \bibnamefont{and} \bibinfo{author}{\bibfnamefont{T.}~\bibnamefont{Hertog}},
  \emph{\bibinfo{title}{{The No-Boundary Measure of the Universe}}}
  (\bibinfo{year}{2007}), \eprint{arXiv:0711.4630 [hep-th]}.

\bibitem[{\citenamefont{Hawking}(2007{\natexlab{b}})}]{Hawking:2007vf}
\bibinfo{author}{\bibfnamefont{S.~W.} \bibnamefont{Hawking}},
  \emph{\bibinfo{title}{{Volume Weighting in the No Boundary Proposal}}}
  (\bibinfo{year}{2007}{\natexlab{b}}), \eprint{arXiv:0710.2029 [hep-th]}.

\bibitem[{\citenamefont{Hartle et~al.}(2008)\citenamefont{Hartle, Hawking, and
  Hertog}}]{Hartle:2008ng}
\bibinfo{author}{\bibfnamefont{J.~B.} \bibnamefont{Hartle}},
  \bibinfo{author}{\bibfnamefont{S.~W.} \bibnamefont{Hawking}},
  \bibnamefont{and} \bibinfo{author}{\bibfnamefont{T.}~\bibnamefont{Hertog}},
  \emph{\bibinfo{title}{{The Classical Universes of the No-Boundary Quantum
  State}}} (\bibinfo{year}{2008}), \eprint{arXiv:0803.1663 [hep-th]}.

\bibitem[{\citenamefont{De~Simone et~al.}(2008)\citenamefont{De~Simone, Guth,
  Salem, and Vilenkin}}]{DeSimone:2008bq}
\bibinfo{author}{\bibfnamefont{A.}~\bibnamefont{De~Simone}},
  \bibinfo{author}{\bibfnamefont{A.~H.} \bibnamefont{Guth}},
  \bibinfo{author}{\bibfnamefont{M.~P.} \bibnamefont{Salem}}, \bibnamefont{and}
  \bibinfo{author}{\bibfnamefont{A.}~\bibnamefont{Vilenkin}},
  \emph{\bibinfo{title}{{Predicting the cosmological constant with the
  scale-factor cutoff measure}}} (\bibinfo{year}{2008}), \eprint{0805.2173}.

\bibitem[{\citenamefont{Winitzki}(2008{\natexlab{b}})}]{Winitzki:2008yb}
\bibinfo{author}{\bibfnamefont{S.}~\bibnamefont{Winitzki}},
  \emph{\bibinfo{title}{{A volume-weighted measure for eternal inflation}}}
  (\bibinfo{year}{2008}{\natexlab{b}}), \eprint{arXiv:0803.1300 [gr-qc]}.

\bibitem[{\citenamefont{Helmer and Winitzki}(2006)}]{Helmer:2006tz}
\bibinfo{author}{\bibfnamefont{F.}~\bibnamefont{Helmer}} \bibnamefont{and}
  \bibinfo{author}{\bibfnamefont{S.}~\bibnamefont{Winitzki}},
  \emph{\bibinfo{title}{Self-reproduction in k-inflation}},
  \bibinfo{journal}{Phys. Rev.} \textbf{\bibinfo{volume}{D74}},
  \bibinfo{pages}{063528} (\bibinfo{year}{2006}), \eprint{gr-qc/0608019}.

\bibitem[{\citenamefont{Tegmark}(2005)}]{Tegmark:2004qd}
\bibinfo{author}{\bibfnamefont{M.}~\bibnamefont{Tegmark}},
  \emph{\bibinfo{title}{What does inflation really predict?}},
  \bibinfo{journal}{JCAP} \textbf{\bibinfo{volume}{0504}}, \bibinfo{pages}{001}
  (\bibinfo{year}{2005}), \eprint{astro-ph/0410281}.

\bibitem[{\citenamefont{Winitzki and Vilenkin}(2000)}]{Winitzki:1999ve}
\bibinfo{author}{\bibfnamefont{S.}~\bibnamefont{Winitzki}} \bibnamefont{and}
  \bibinfo{author}{\bibfnamefont{A.}~\bibnamefont{Vilenkin}},
  \emph{\bibinfo{title}{Effective noise in stochastic description of
  inflation}}, \bibinfo{journal}{Phys. Rev.} \textbf{\bibinfo{volume}{D61}},
  \bibinfo{pages}{084008} (\bibinfo{year}{2000}), \eprint{gr-qc/9911029}.

\bibitem[{\citenamefont{Vilenkin}(1983)}]{Vilenkin:1983xq}
\bibinfo{author}{\bibfnamefont{A.}~\bibnamefont{Vilenkin}},
  \emph{\bibinfo{title}{The birth of inflationary universes}},
  \bibinfo{journal}{Phys. Rev.} \textbf{\bibinfo{volume}{D27}},
  \bibinfo{pages}{2848} (\bibinfo{year}{1983}).

\bibitem[{\citenamefont{Starobinsky}(1986)}]{Starobinsky:1986fx}
\bibinfo{author}{\bibfnamefont{A.~A.} \bibnamefont{Starobinsky}},
  \emph{\bibinfo{title}{{Stochastic de Sitter (inflationary) stage in the early
  universe}}} (\bibinfo{year}{1986}), \bibinfo{note}{in: Current Topics in
  Field Theory, Quantum Gravity and Strings, Lecture Notes in Physics 206, eds.
  H.J. de Vega and N. Sanchez (Springer Verlag), p. 107}.

\bibitem[{\citenamefont{Goncharov et~al.}(1987)\citenamefont{Goncharov, Linde,
  and Mukhanov}}]{Goncharov:1987ir}
\bibinfo{author}{\bibfnamefont{A.~S.} \bibnamefont{Goncharov}},
  \bibinfo{author}{\bibfnamefont{A.~D.} \bibnamefont{Linde}}, \bibnamefont{and}
  \bibinfo{author}{\bibfnamefont{V.~F.} \bibnamefont{Mukhanov}},
  \emph{\bibinfo{title}{The global structure of the inflationary universe}},
  \bibinfo{journal}{Int. J. Mod. Phys.} \textbf{\bibinfo{volume}{A2}},
  \bibinfo{pages}{561} (\bibinfo{year}{1987}).

\bibitem[{\citenamefont{Stakgold}(1979)}]{Stakgold:1979:GF}
\bibinfo{author}{\bibfnamefont{I.}~\bibnamefont{Stakgold}},
  \emph{\bibinfo{title}{Green's functions and boundary value problems}}
  (\bibinfo{publisher}{Wiley}, \bibinfo{address}{New York},
  \bibinfo{year}{1979}).

\bibitem[{\citenamefont{Podolsky et~al.}(2008)\citenamefont{Podolsky, Majumder,
  and Jokela}}]{Podolsky:2008du}
\bibinfo{author}{\bibfnamefont{D.~I.} \bibnamefont{Podolsky}},
  \bibinfo{author}{\bibfnamefont{J.}~\bibnamefont{Majumder}}, \bibnamefont{and}
  \bibinfo{author}{\bibfnamefont{N.}~\bibnamefont{Jokela}},
  \emph{\bibinfo{title}{{Disorder on the landscape}}} (\bibinfo{year}{2008}),
  \eprint{0804.2263}.

\end{thebibliography}

\end{document}